\documentclass{elsart3-1}

\usepackage{graphicx}
\usepackage{longtable}
\DeclareGraphicsExtensions{.png,.jpg,.pdf}

\usepackage{amssymb}

\usepackage[english,francais]{babel}

\newtheorem{e-proposition}[theorem]{Proposition}

\newtheorem{e-definition}[theorem]{Definition\rm}

\renewcommand{\theequation}{\arabic{equation}}
\def\og{\leavevmode\raise.3ex\hbox{$\scriptscriptstyle\langle\!\langle$~}}
\def\fg{\leavevmode\raise.3ex\hbox{~$\!\scriptscriptstyle\,\rangle\!\rangle$}}

\begin{document}
\centerline{Physics in high magnetic field/Physique en champ magn\'{e}tique intense}
\begin{frontmatter}


\selectlanguage{english}
\title{Heavy fermions in high magnetic field}


\selectlanguage{english}
\author[authorlabel1]{D.~Aoki},
\ead{dai.aoki@cea.fr}
\author[authorlabel2]{W.~Knafo},
\ead{william.knafo@lncmi.cnrs.fr}
\author[authorlabel3]{I.~Sheikin},
\ead{ilya.sheikin@lncmi.cnrs.fr}

\address[authorlabel1]{Institut Nanosciences et Cryog\'{e}nie, SPSMS, CEA-Grenoble, 17 rue des Martyrs, 38054 Grenoble, France}
\address[authorlabel2]{Laboratoire National des Champs Magn\'{e}tiques Intenses, UPR 3228, CNRS-UJF-UPS-INSA, 143 Avenue de Rangueil,
31400 Toulouse, France.}
\address[authorlabel3]{Laboratoire National des Champs Magn\'{e}tiques Intenses, UPR 3228, CNRS-UJF-UPS-INSA, 25 rue des Martyrs, B.P. 166,
38042 Grenoble cedex 9, France.}

\begin{abstract}
We give an overview on experimental studies performed in the last 25 years on heavy-fermion systems in high magnetic field. The properties of
field-induced magnetic transitions in heavy-fermion materials close to a quantum antiferromagnetic-to-paramagnetic instability are presented. Effects
of a high magnetic field to the Fermi surface, in particular the splitting of spin-up and spin-down bands, are also considered. Finally, we review on
recent advances on the study of non-centrosymmetric compounds and ferromagnetic superconductors in a high magnetic field.

{\it To cite this article: D.~Aoki, W.~Knafo, and I.~Sheikin, C. R. Physique XX (2012).}

\vskip 0.5\baselineskip

\selectlanguage{francais} \noindent{\bf R\'esum\'e} \vskip 0.5\baselineskip \noindent {\bf Fermions lourds en champ magn\'{e}tique intense.} Cette
revue donne un aper\c{c}u des \'{e}tudes exp\'{e}rimentales faites depuis 25 ans sur les syst\`{e}mes \`{a} fermions lourds en champ magn\'{e}tique
intense. Les propri\'{e}t\'{e}s des transitions de phase magn\'{e}tiques de compos\'{e}s proches d'une instabilit\'{e} antiferromagn\'{e}tique sont
pr\'{e}sent\'{e}es. Les effets d'un champ magn\'{e}tique intense sur la surface de Fermi, en particulier la s\'{e}paration des bandes de spin "up" et
"down", sont aussi consid\'{e}r\'{e}es. Finalement, nous faisons le point sur les avanc\'{e}es r\'{e}centes dans l'\'{e}tude en champ magn\'{e}tique
intense de compos\'{e}s non-centrosymm\'{e}triques et de supraconducteurs ferromagn\'{e}tiques.

{\it Pour citer cet article~: D.~Aoki, W.~Knafo, and I.~Sheikin, C. R. Physique XX (2012).}

 \vskip 0.5\baselineskip
\noindent{\small{\it Mots-cl\'es~:} Fermions lourds; Champ magn\'{e}tique intense; M\'{e}tamagn\'{e}tisme, Surface de Fermi; Supraconducteurs
ferromagn\'{e}tiques}}
\end{abstract}
\end{frontmatter}

\selectlanguage{english}
\section{Introduction}\label{intro}

Heavy-fermion systems \cite{Flouquet2005,Lohneysen2007} are intermetallic materials composed of rare earths (Ce, Yb) or actinides (U, Np, Pt)
elements. In these systems, partially filled \(4f\)- or \(5f\)-electron orbitals are strongly-coupled to conduction-electrons bands. Electronic
interactions give rise to the formation of heavy quasiparticles, i.e., narrow electronic bands with a strong enhancement of the effective mass $m^*$,
which typically reaches 100 to 1000 times the value of the free-electron mass $m_0$. The magnetic properties of heavy fermions are governed by a
subtle competition between Kondo and Ruderman-Kittel-Kasuya-Yosida (RKKY) interactions, both of which depending on the product of the exchange
interaction \(J\) between a $f$-electron and a conduction electron by the density \(n(E_F)\) of states at the Fermi level. The Kondo interaction can
be seen as a screening of $f$-electron magnetic moments by the conduction electrons, which tends to destabilize the magnetic ordering of the
$f$-electron moments. Its characteristic energy scale, the Kondo temperature, can be approximated by:
\begin{equation}\label{Kondo}
T_K \propto \exp(-1/J n(E_F)).
\end{equation}
The RKKY interaction is a magnetic exchange interaction between the $f$-electron moments mediated by the conduction electrons, which promotes
long-range magnetic ordering. Its energy scale, the RKKY temperature, can be approximated by:
\begin{equation}\label{RKKY}
T_{RKKY} \propto J^2 n(E_F).
\end{equation}

\begin{figure}[b]
\centering
\includegraphics[width=0.5\textwidth]{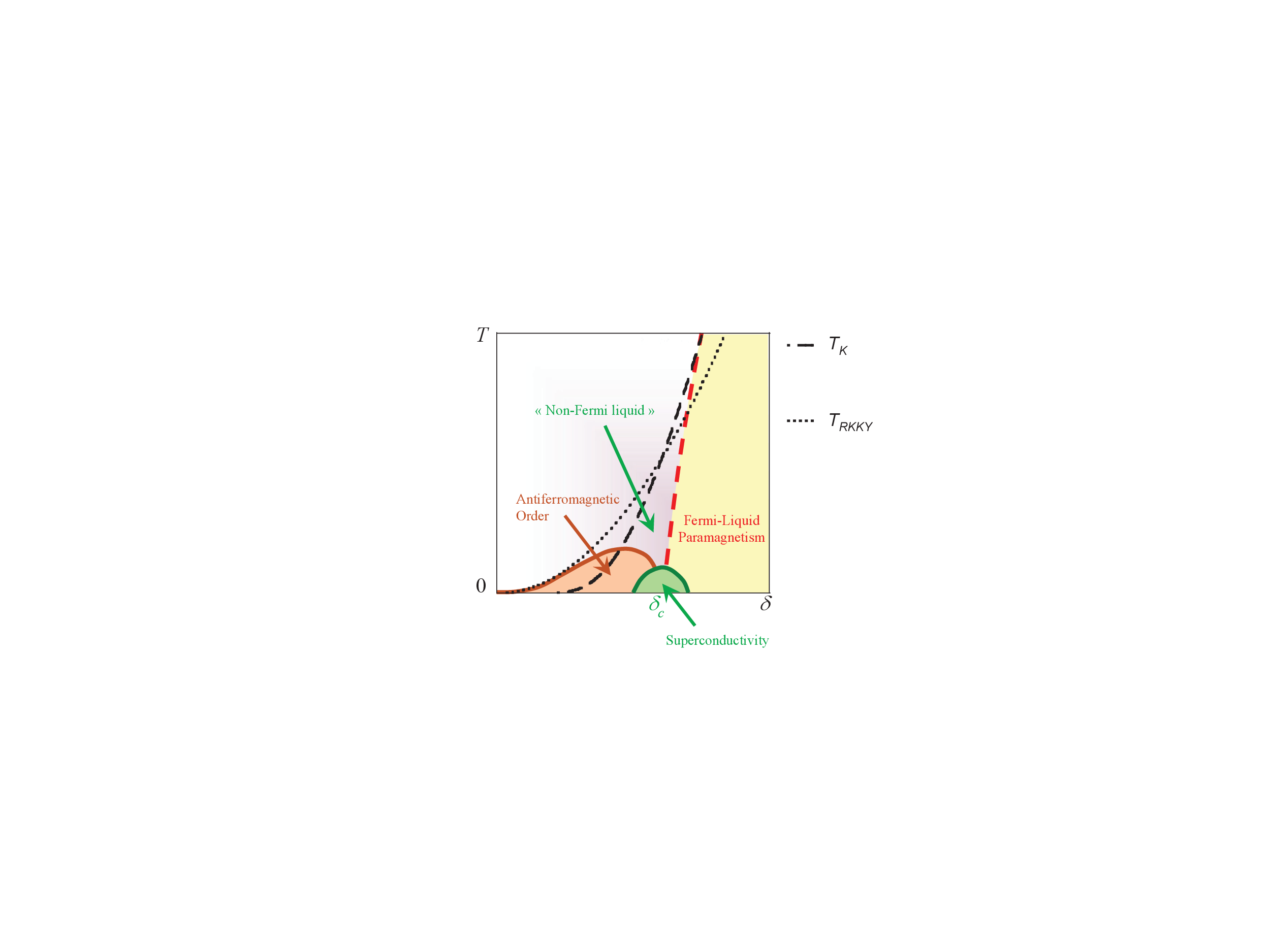}
\caption{Doniach's phase diagram. The short dashed line is the RKKY temperature, $T_{RKKY}$, at which magnetic order would occur in the absence of
Kondo screening. The long dashed line is the Kondo temperature, $T_K$, below which the $f$-electron moments are screened by the conduction electrons.
When $T_{RKKY} = T_K$, the magnetic ordering temperature is driven to zero giving rise to a quantum phase transition between an antiferromagnetic and
a paramagnetic Fermi liquid ground states. In many heavy-fermion systems, a superconducting pocket develops in the vicinity of the quantum phase
transition.} \label{Doniach}
\end{figure}

The parameter $\delta=Jn(E_F)$ can be varied by applying either hydrostatic or chemical pressure. In many systems, it is possible to drive by
pressure or doping the magnetic ordering temperature to zero, giving rise to a quantum phase transition (in the limit of zero-temperature) between an
antiferromagnetic and a paramagnetic ground states. The Doniach's phase diagram~\cite{Doniach1977,Iglesias1997} drawn in figure~\ref{Doniach}
schematically illustrates how the competition of Kondo and RKKY interactions can lead to an antiferromagnetic-to-paramagnetic quantum instability. In
the vicinity of the quantum phase transition, the amplitude of quantum magnetic fluctuations grows rapidly, leading to two distinct phenomena:
non-Fermi liquid behavior~\cite{Stewart2001} and unconventional superconductivity \cite{Pfleiderer2009}. A non-Fermi liquid is a strong deviation
from a Fermi-liquid behavior at low temperature, observed in resistivity [$\rho(T) = \rho_0 + AT^n$, $1<n<1.5$ instead of $\rho(T) = \rho_0 + AT^2$],
magnetic susceptibility [sometimes $\chi(T) \propto \chi_0(1-c(T/T_0)^{1/2})$ instead of $\chi(T) = \chi_0(1+aT^2)$] and specific heat (sometimes
$C/T \propto -\ln T$ instead of $C/T = \gamma + \beta T^2$) measurements. Interplay between magnetism and superconductivity is at the heart of the
heavy-fermion problem, since unconventional superconductivity often develops in the vicinity of a quantum phase transition~\cite{Pfleiderer2009},
being only observed in very clean heavy-fermion samples. Electron pairing is probably due to quantum critical magnetic fluctuations (instead of
phonons as in conventional superconductors), whose intensity is maximal at the quantum phase transition. The question of the nature of quantum
critical magnetic fluctuations is still a matter of strong debate. A scenario based on critical fluctuations of the magnetic order parameter has
initially been proposed \cite{Moriya1995,Hertz1976,Millis1993}, but it failed to describe the non-Fermi liquid regime. An unconventional scenario
based on local, i.e., wavevector $\mathbf{q}$-independent, critical magnetic fluctuations has also been proposed by some authors \cite{Coleman2001}.
More recently, an enhancement of the antiferromagnetic fluctuations, but not of the local magnetic fluctuations, was observed at the quantum phase
transition in the heavy-fermion system Ce$_{1-x}$La$_x$Ru$_2$Si$_2$ ~\cite{Knafo2009} permitting to rehabilitate, at least for this system, the
conventional scenario based on quantum critical fluctuations of the order-parameter. Another issue is to determine whether the \(f\)-electrons are
itinerant or localized on both sides of the quantum phase transition. This can be achieved by comparing the experimentally-obtained angular
dependence of de Haas-van Alphen frequencies with the results of theoretical band-structure calculations performed for both itinerant and localized
\(f\)-electrons. de Haas-van Alphen measurements also provide the effective masses for each band, whose behavior on crossing a quantum phase
transition can be investigated.

\begin{figure}[t]
\centering
\includegraphics[width=.9\textwidth]{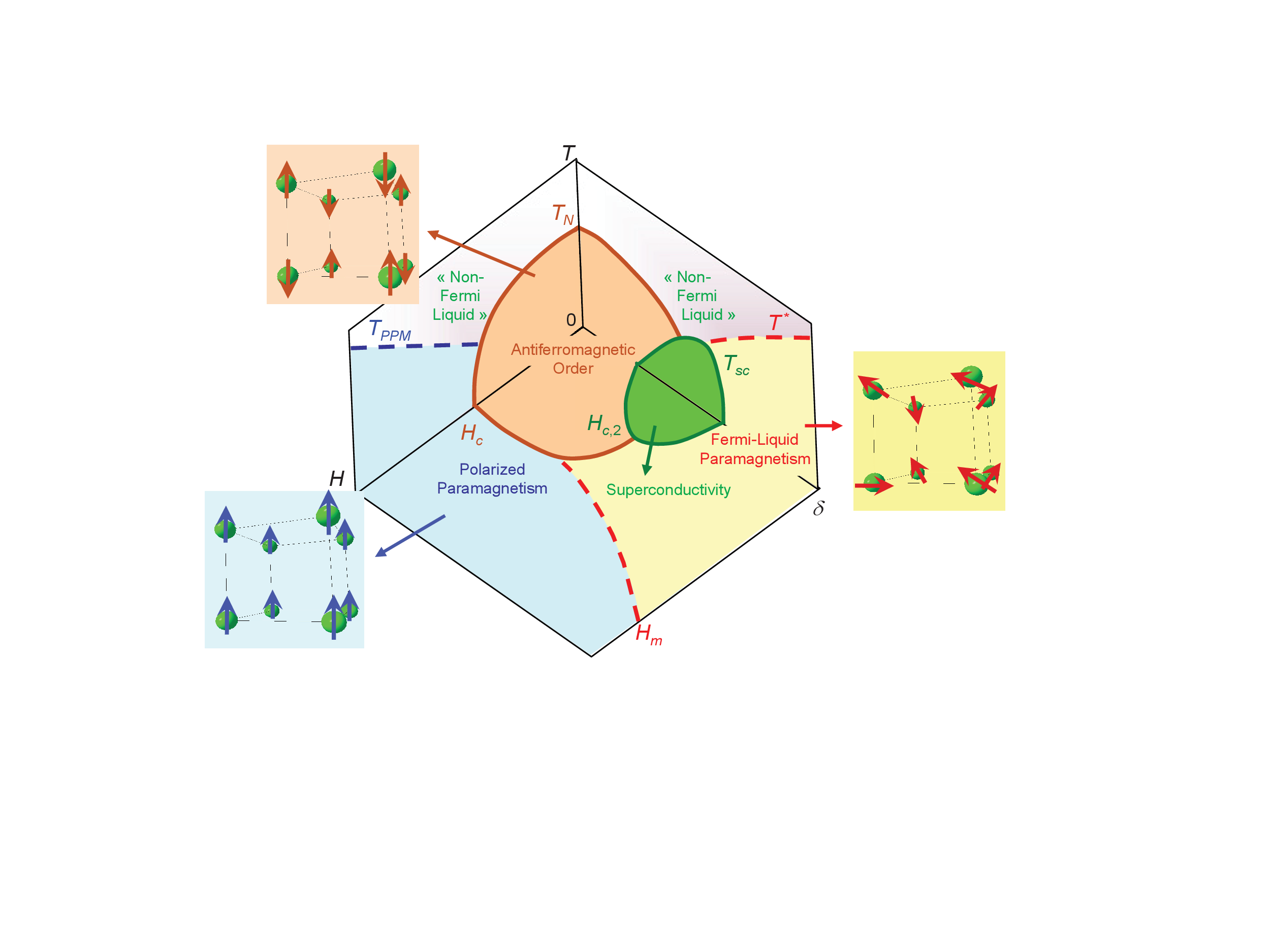}
\caption{Schematic three-dimensional ($T,\delta,H$) phase diagram of heavy-fermion systems close to an antiferromagnetic-to-paramagnetic quantum
instability, where $\delta$ can be pressure or doping.} \label{3D}
\end{figure}

While many heavy fermions have been initially studied under pressure and/or doping, these last years have shown the emergence of numerous works under
a magnetic field $\mathbf{H}$. The three main phenomena associated with pressure- or doping-induced quantum criticality have been already observed in
a magnetic field, i.e. i) field-induced non-Fermi liquid (e.g. in YbRh$_2$Si$_2$~\cite{Custers2003}), ii) field-induced quantum critical fluctuations
(of ferromagnetic nature in CeRu$_2$Si$_2$~\cite{Sato2001,Raymond1999,Flouquet2004}, and iii) field-induced superconductivity (in URhGe
\cite{Levy2005,Levy2007}). Contrary to pressure and doping, a magnetic field has the advantage that it can be varied continuously. Under a magnetic
field, heavy-fermion systems undergo a transition to a polarized paramagnetic regime. The three-dimensional phase diagram ($\delta,H,T$) shown in
Fig. \ref{3D} illustrates schematically how the polarized regime is reached from the zero-field antiferromagnetic or paramagnetic ground states. The
present review aims to give an overview of recent state-of-art studies of heavy fermions in a magnetic field, with a particular focus on experiments
performed in very high fields above 20~T. Section \ref{MM} focuses on field-induced quantum criticality in heavy-fermion systems close to an
antiferromagnetic instability. The magnetic and Fermi surface properties at a quantum phase transition under magnetic field will be considered for
heavy-fermion antiferromagnets and paramagnets, including the exotic 'hidden-order' material URu$_2$Si$_2$. Magnetic field is expected to modify the
magnetic correlations and, therefore, the electronic structure of heavy-fermion compounds, i.e., the Fermi surface topology and effective masses. The
study of heavy-fermion Fermi surfaces in high-field is considered in detail in Section \ref{effective_mass}, where the splitting by a magnetic field
of the Fermi surface into majority-spin and minority-spin surfaces, with different spin-up and spin-down effective masses, as predicted by most
theoretical models, is considered. Following the discovery of superconductivity in the non-centrosymmetric heavy-fermion compound CePt$_3$Si
\cite{Bauer2004}, materials without inversion symmetry in their crystal structure have attracted a lot of experimental and theoretical interest. The
interest is mainly due to a fascinating theoretical prediction~\cite{Edelshtein1989,Gorkov2001} that superconductive pairing in such systems requires
an admixture of a spin-singlet with a spin-triplet state. Very unusual superconducting properties of these materials, such as enormously high upper
critical field, will be discussed in Section~\ref{Non-centrosymmetric}. Finally, Section \ref{FM} synthesizes the high-field properties of the
recently-discovered class of ferromagnetic superconductors. This family is composed of UGe$_2$~\cite{Saxena2000}, URhGe~\cite{Aoki2001}, UIr
\cite{Akazawa2004}, and UCoGe~\cite{Huy2007}. While UGe$_2$ and UIr become superconducting under pressure, URhGe and UCoGe are superconducting at
ambient pressure, well below their Curie temperature. Field-induced superconductivity develops in URhGe around a field of 12~T applied along
$\mathbf{b}$, which is well above the critical superconducting field of 2.5~T \cite{Levy2005}, and results from the presence of a field-induced
magnetic transition.

\section{Field-induced magnetic transitions}\label{MM}

\begin{figure}[b]
\centering
\includegraphics[width=.9\textwidth]{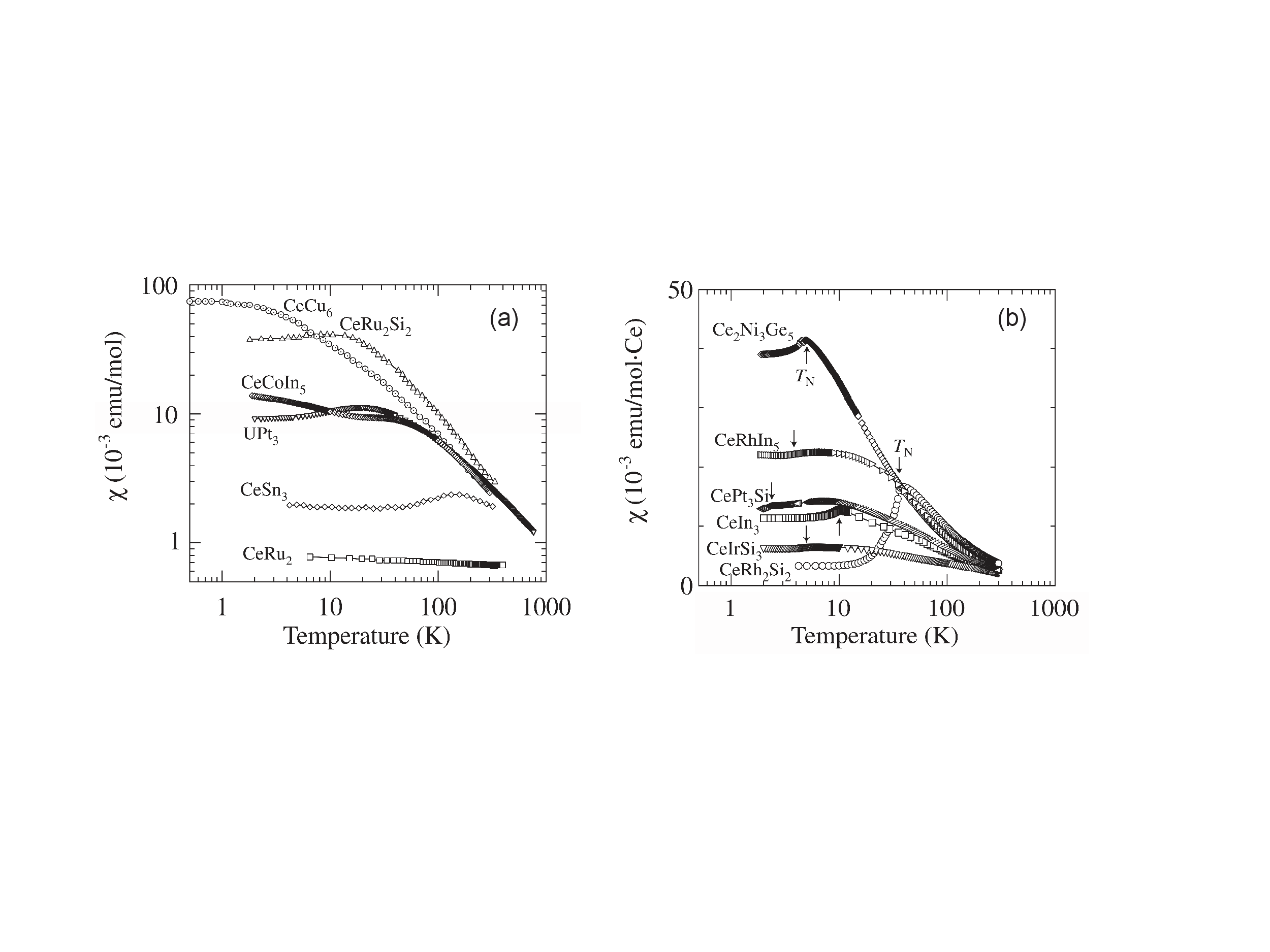}
\caption{Magnetic susceptibility of various heavy-fermion (a) paramagnets and (b) antiferromagnets \cite{Settai2007}}
\label{susc}
\end{figure}

Field-induced magnetic transitions have been observed in many heavy-fermion materials with either a paramagnetic or a magnetically-ordered ground
state. Among the former are, for example, the prototypical CeRu\(_2\)Si\(_2\)~\cite{Haen1987,Flouquet2002} and
CeCu\(_6\)~\cite{Schroder1992,Lohneysen1993} heavy-fermion paramagnets. The latter include CeRh\(_2\)Si\(_2\)~\cite{Knafo2010}
CePd\(_2\)Si\(_2\)~\cite{Sheikin2003}, and YbRh\(_2\)Si\(_2\)~\cite{Tokiwa2004}. In many systems, a strong magnetic anisotropy leads at low
temperature to field-induced first-order transitions. Such transitions, called metamagnetic transitions, are characterized by a sudden step-like
increase of the magnetization at the onset of polarized paramagnetism \cite{Stryjewski1977}. As pointed out for itinerant magnets
\cite{Sakakibara1990,Sakakibara1990b} and heavy-fermion magnets \cite{Inoue2001,Fukuhara1996,Sugiyama2002,Onuki2004,Takeuchi2010}, which are a
sub-class of itinerant magnets, the transition field is often related to the temperature $T_{\chi,max}$ at the maximum of the magnetic susceptibility
by a correspondence 1~T~$\leftrightarrow$~1~K. From the Maxwell relation $(\partial M/\partial T)_H=(\partial S/\partial H)_T$, the magnetic entropy
$S$ and thus the magnetic fluctuations increase with the magnetic field $H$ when the temperature is smaller than $T_{\chi,max}$, indicating the
proximity of a field-induced transition \cite{BealMonod1982}. A maximum in the magnetic susceptibility at low magnetic field, as shown in Figure
\ref{susc} for several heavy-fermion paramagnets and antiferromagnets \cite{Settai2007}, is thus generally the first indication for a high-field
transition. A metamagnetic transition can also be driven to zero temperature~\cite{Millis2002} by tuning an additional parameter, leading to a
quantum critical end-point (cf. the bilayer ruthenate Sr\(_3\)Ru\(_2\)O\(_7\)~\cite{Grigera2001,Grigera2004}, where the field-tuned quantum critical
point also leads to new -and not yet identified- low-temperature phases). In this Section we present an overview of field-induced magnetic
transitions in heavy-fermion materials close to a paramagnetic-to-antiferromagnetic instability (the case of ferromagnetic heavy fermions will be
considered separately in Section \ref{FM}). Sub-section \ref{pm} presents the high-field properties of paramagnetic systems and Sub-section \ref{af}
is devoted to heavy-fermion antiferromagnets. Sub-section \ref{fsm} shows that Fermi surface modifications can be induced at a heavy-fermion
field-induced transition. Finally, Sub-section \ref{ho} focuses on the exotic case of URu$_2$Si$_2$ where the hidden-order state, whose unknown
nature is still a matter of strong debate, can be destabilized by a high magnetic field.

\subsection{Heavy-fermion paramagnets}\label{pm}

\begin{table}[b]
\begin{center}
\caption{Temperature of the maximum (or kink) in the magnetic susceptibility, magnetic field of the transition, field direction, and inelastic
neutron scattering linewidth in various heavy-fermion paramagnets.}
\begin{tabular}{lccccc}
Material&$T_{\chi,max}$ (K)&$H_m$ (T)&$\mathbf{H}\parallel$&Linewidth $\Gamma$ (K)&References\\
\hline
CeRu$_2$Si$_2$&10&7.8&$\mathbf{c}$&10 $(\mathbf{k}_1,T\rightarrow 0)$&\cite{Fisher1991,Paulsen1990,Raymond2007,Knafo2009}\\
CeRu$_2$Si$_2$ ($p=1\rightarrow2$~kbar)&11.7 $\rightarrow$ 14&10.3 $\rightarrow$ 13&$\mathbf{c}$&-&\cite{Mignot1988}\\
Ce$_{1-x}$Y$_{x}$Ru$_2$Si$_2$ ($x=1.5\rightarrow10$~\%)&12.5 $\rightarrow$ 25&9.5 $\rightarrow$ 19.3&$\mathbf{c}$&-&\cite{Park1995,Haen1995}\\
Ce$_{1-x}$La$_{x}$Ru$_2$Si$_2$ ($x=3\rightarrow7.5$~\%)&7.4 $\rightarrow$ 5&6.2 $\rightarrow$ 4&$\mathbf{c}$&- $\rightarrow$ 2.5 $(\mathbf{k}_1,T\rightarrow 0)$&\cite{Fisher1991,Matsumoto2008,Matsumoto2010,Knafo2004}\\
CeRu$_2$(Si$_{1-x}$Ge$_{x}$)$_2$ ($x=3.5\rightarrow7$~\%)&6 $\rightarrow$ 3.8&6 $\rightarrow$ 4.2&$\mathbf{c}$&-&\cite{Matsumoto2011}\\
CeFe$_2$Ge$_2$&(kink at $\simeq20$~K)&30&$\mathbf{c}$&-&\cite{Sugawara1999,Ebihara1995}\\
CeNi$_2$Ge$_2$&28&42 (powder)&$\mathbf{c}$&40 $(\mathbf{q}_0,T\rightarrow 0)$&\cite{Fukuhara1996,Fak2000}\\
CeIrIn$_5$&(kink at $\simeq30$~K)&30-40&$\mathbf{c}$&30 ($\langle\mathbf{Q}\rangle$,$T=8$~K)&\cite{Palm2003,Takeuchi2001,Willers2010}\\
CeCu$_6$&(kink at $\simeq1-1.5$~K)&1.7&$\mathbf{c}$&2.5 $((1.15,0,0),T\rightarrow 0)$&\cite{Schroder1992,Lohneysen1993,RossatMignod1988}\\
CeTiGe&25&12.5&(polycrystal)&-&\cite{Deppe2012}\\
CeFePO&5&4&$\mathbf{c}$&-&\cite{Kitagawa2011}\\
YbIr$_2$Zn$_{20}$&7.4&9.7 (12)&$[100]$ ($[110]$)&-&\cite{Takeuchi2010}\\
YbRh$_2$Zn$_{20}$&5.3&6.4&$[100]$&-&\cite{Hirose2011}\\
YbCo$_2$Zn$_{20}$&0.32&0.6 (0.57)&$[100]$ ($[110]$)&-&\cite{Hirose2011,Takeuchi2011}\\
YbCu$_{5-x}$Ag$_{x}$ ($x=0\rightarrow1$)&17 $\rightarrow$ 40&10 $\rightarrow$ 40&(cubic)&-&\cite{Sarrao1999,Tsujii2001}\\
USn$_3$&(kink at $\simeq15-20$~K)&30&$\mathbf{a}$&60 ($\langle\mathbf{Q}\rangle$)&\cite{Sugiyama2002,Loewenhaupt1990}\\
URu$_2$Si$_2$&55&35-39&$\mathbf{c}$&50 $(\mathbf{Q}_1,T\geq T_0)$&\cite{Inoue2001,Scheerer2012,Broholm1991}\\
\end{tabular}
\label{tablePM}
\end{center}
\end{table}

\begin{figure}[t]
\centering
\includegraphics[width=0.6\textwidth]{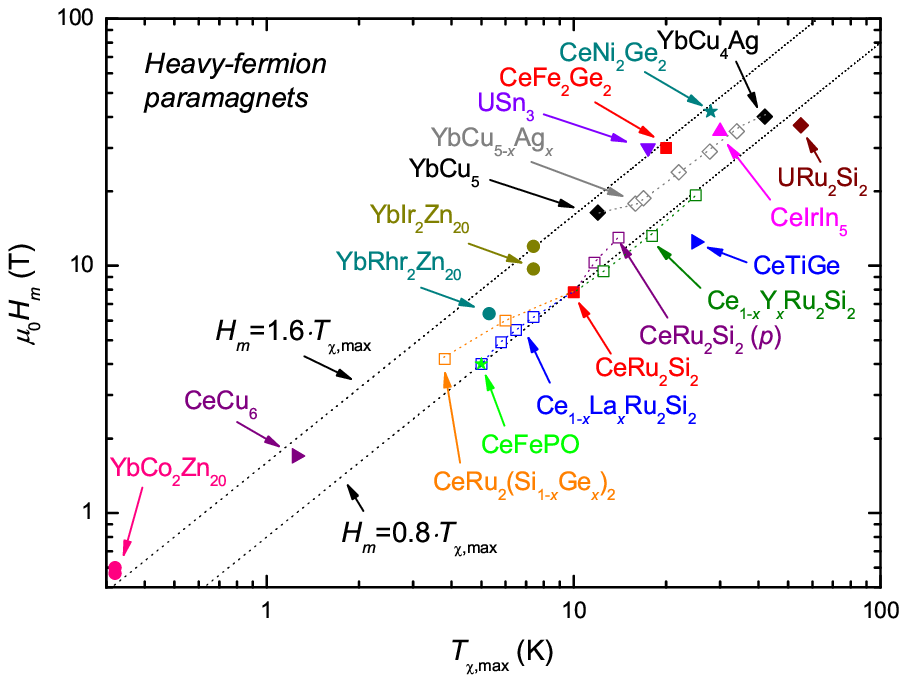}
\caption{Metamagnetic field as a function of the temperature at the maximum (or kink) of the magnetic susceptibility for various heavy-fermion
paramagnets
\cite{Schroder1992,Lohneysen1993,Inoue2001,Fukuhara1996,Fisher1991,Paulsen1990,Sugawara1999,Ebihara1995,Palm2003,Takeuchi2001,Deppe2012,Scheerer2012,Mignot1988,Park1995,Haen1995,Sugiyama1999,Sugiyama2002}.}
\label{BTchimax}
\end{figure}

Table \ref{tablePM} summarizes the characteristics of several heavy-fermion paramagnets undergoing a field-induced transition to a polarized
paramagnetic regime. In this Table, the metamagnetic (called sometimes pseudo-metamagnetic when the transition is not clearly first-order) field
$H_m$, the temperature $T_{\chi,max}$ at the maximum of the susceptibility, and the corresponding field direction are given for each system. Figure
\ref{BTchimax} presents in a log-log scale a $H_m$ versus $T_{\chi,max}$ plot of these data. It extends to a higher number ($\simeq30$) of
heavy-fermion materials the similar plots already presented in Refs. \cite{Inoue2001,Fukuhara1996,Sugiyama2002,Onuki2004,Takeuchi2010,Hirose2011},
and is restricted here to paramagnets. A $H_c$ versus $T_N$ plot will be shown for heavy-fermion antiferromagnets in Sub-section \ref{af}. A striking
feature of Figure \ref{BTchimax} is that $H_m$ and $T_{\chi,max}$ are almost connected by a simple correspondence 1~T~$\leftrightarrow$~1~K. Most of
the compounds considered are such that $0.8\cdot T_{\chi,max}<H_m<1.6\cdot T_{\chi,max}$, despite $H_m$ and $T_{\chi,max}$ vary by more than two
decades, from 0.57~T and 0.32~K, respectively, in YbCo$_2$Zn$_{20}$ ($\mathbf{H}\parallel[110]$) to 35-39~T and 55~K, respectively, in URu$_2$Si$_2$
($\mathbf{H}\parallel\mathbf{c}$). The fact that $H_m$ and $T_{\chi,max}$ are almost linearly connected indicates that they are mainly controlled by
a single energy scale. In the literature (see for example Refs. \cite{Inoue2001,Sugiyama2002,Onuki2004,Takeuchi2010,Hirose2011}), it has often been
proposed that $T_{\chi,max}$ in heavy-fermion paramagnets is related to the onset of the Kondo hybridization between $f$- and conduction electrons,
i.e., to a crossover from a high-temperature regime where the $f$-electrons are localized to a low-temperature regime where they are itinerant. This
hypothesis is compatible with the assumption proposed in Ref. \cite{Kitagawa2011} that the metamagnetic field $H_m$ in CeFePO is associated with a
breakdown of the Kondo effect. These hypotheses assume that $T_{\chi,max}$ and $H_m$ are respectively related to the temperature- and magnetic
field-induced destruction of the Kondo hybridization, as expected from a single-impurity Kondo model \cite{Rajan1982,Rajan1983}. They also imply that
a drastic reduction of $\mathbf{q}$-independent Kondo magnetic fluctuations \cite{Knafo2009} should occur for $T>T_{\chi,max}$ and $H>H_m$,
respectively, which has never been shown experimentally yet. Other theories have been proposed to describe metamagnetism in heavy-fermion
paramagnets, from single-site models \cite{Wohlfarth1962,Evans1992,Yamada1993,Hanzawa1997,Ono1998} to models where intersite correlations are also
considered \cite{Konno1991,Takahashi1998,Satoh1998}. We show below that neutron experiments on the heavy-fermion paramagnet CeRu$_2$Si$_2$ indicate
that $T_{\chi,max}$ and $H_m$ in this system are related to intersite antiferromagnetic correlations, but not to the Kondo effect.

\begin{figure}[t]
\centering
\includegraphics[width=.9\textwidth]{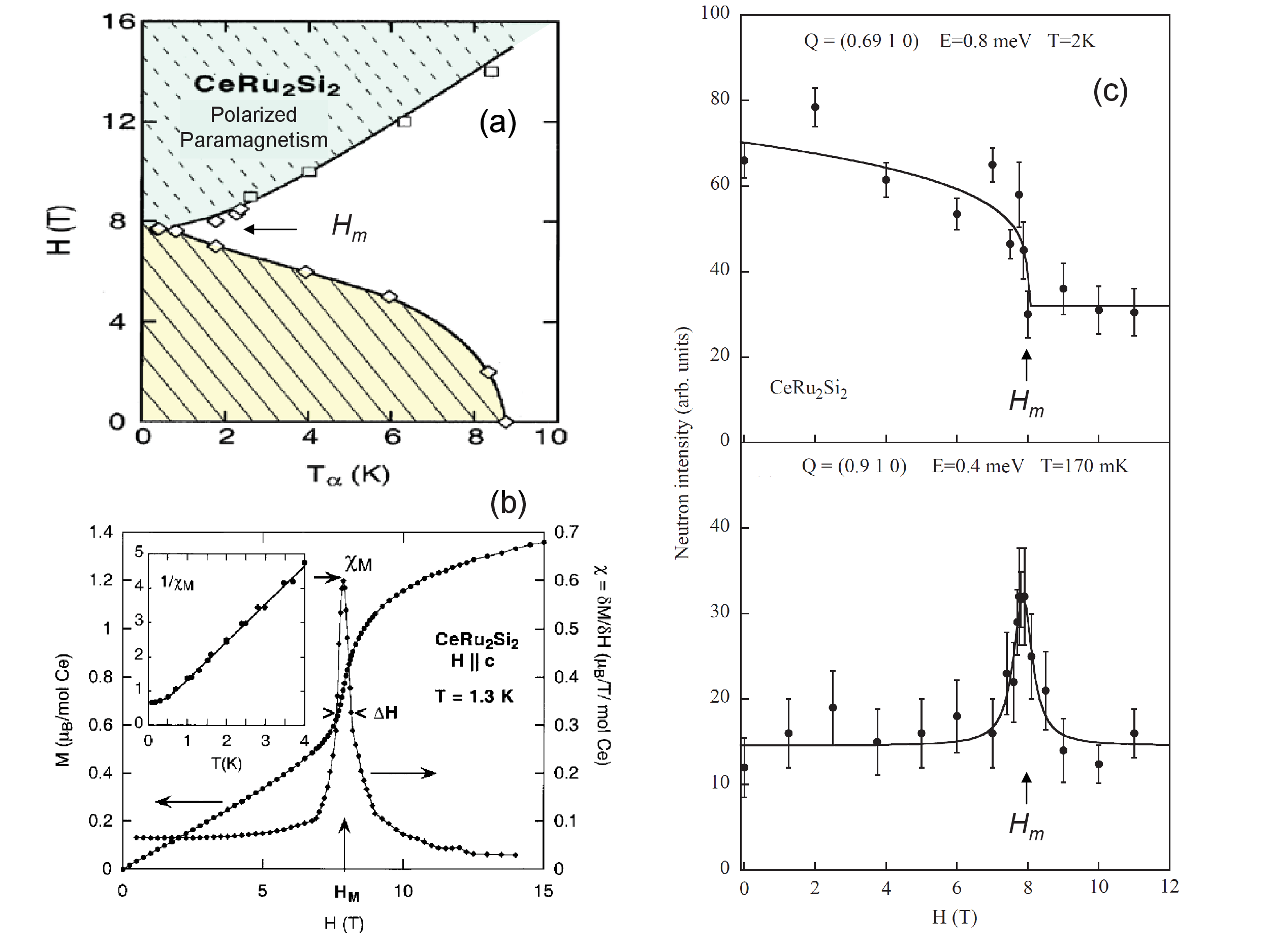}
\caption{(a) ($T,\delta,H$) phase diagram probed by thermal expansion and heat capacity \cite{Flouquet2002}, (b) magnetization versus field
\cite{Flouquet2002}, and (c) inelastic neutron scattering intensity at wavevectors characteristic of the antiferromagnetic (top) and ferromagnetic
(bottom) fluctuations \cite{Flouquet2004}, of CeRu$_2$Si$_2$ with $\mathbf{H}\parallel\mathbf{c}$.} \label{cerusi}
\end{figure}

CeRu$_2$Si$_2$ has probably been the mostly-studied heavy-fermion system since twenty years. The proximity of this paramagnet to antiferromagnetic
instabilities is demonstrated by the increase on cooling of the specific heat divided by temperature $C_p/T$ \cite{Fisher1991} and of the electronic
Gr\"{u}neisen parameter \cite{Lacerda1989}. RKKY interactions lead to maxima of the dynamical magnetic susceptibility at the incommensurate wave
vectors $\mathbf{k}_1=(0.31,0,0)$, $\mathbf{k}_2=(0.31,0.31,0)$, and $\mathbf{k}_3=(0,0,0.35)$, as probed by inelastic neutron scattering
\cite{Kadowaki2004}. At wave vectors sufficiently far from $\mathbf{k}_1$, $\mathbf{k}_2$, and $\mathbf{k}_3$ magnetic fluctuations persist and can
be considered as the signature of a local (or single-site) Kondo effect. Chemical doping (with La, Ge, or Rh) and/or magnetic field tunings can favor
one particular interaction and establish antiferromagnetic long-range ordering with either $\mathbf{k}_1$, $\mathbf{k}_2$, or $\mathbf{k}_3$ wave
vectors \cite{Quezel1988,Haen2002,Mignot1991,Watanabe2003}. Upon doping by La, it has been shown that the quantum phase transition to an
antiferromagnetic state is governed by fluctuations of the antiferromagnetic order parameter (moment with the wavevector $\mathbf{k}_1$)
\cite{Knafo2009}. Under a magnetic field applied along the easy axis $\mathbf{c}$, a magnetic transition to a polarized paramagnetic regime occurs at
$H_m=7.8$~T \cite{Paulsen1990} [Figure \ref{cerusi}(a-b)]. As well as the temperature $T_{\alpha,max}$ at the maximum of thermal expansion
\cite{Paulsen1990}, the temperature $T_{\chi,max}$ at the maximum of susceptibility \cite{Ishida1998} decreases in a magnetic field
$\mathbf{H}\parallel\mathbf{c}$ and vanishes above $H_m$. $T_{\chi,max}$ and $T_{\alpha,max}$ are related to the same phenomenon and they delimitate
a low-temperature regime which vanishes above $H_m$. The correspondence 1~K~$\leftrightarrow$~1~T between the maximum of susceptibility
$T_{\chi,max}=10$~K and the critical magnetic field $H_m$ in CeRu$_2$Si$_2$ suggests that both $T_{\chi,max}$ and $H_m$ are controlled by a single
magnetic energy scale. Antiferromagnetic fluctuations at $\mathbf{k}_1$ vanish at $H_m$, while ferromagnetic fluctuations develop in a narrow window
around $H_m$ \cite{Sato2001,Raymond1999,Flouquet2004} [Figure \ref{cerusi}(c)]. At zero-field, antiferromagnetic fluctuations with $\mathbf{k}_1$
saturate below $\simeq10$~K and their temperature scale is given by the linewidth $\Gamma(\mathbf{k}_1)=10$~K~$=T_{\chi,max}$
\cite{Knafo2009,Raymond2007}. $T_{\chi,max}$ and $H_m$ correspond thus to the energy scale of the antiferromagnetic fluctuations with the wavevector
$\mathbf{k}_1=(0.31,0,0)$ and the regime below $T_{\chi,max}$ and $H_m$ is controlled by these intersite magnetic fluctuations. When crossing
$T_{\chi,max}$ or $H_m$, no drastic change of local ($\mathbf{q}$-independent) Kondo magnetic fluctuations has been observed
\cite{Knafo2009,Raymond1999,Raymond2007}, which indicates no sudden modification of the single-site Kondo effect and thus no Kondo breakdown at
$T_{\chi,max}$ or $H_m$.

CeRu$_2$Si$_2$ is the unique heavy-fermion compound which has been investigated on both sides of its metamagnetic transition by a full set of
experimental techniques made of electronic transport, thermodynamic, elastic and inelastic neutron scattering, de Haas-van Alphen, and x-rays
absorption spectroscopy experiments (see also Sub-section \ref{fsm}). Contrary to CeRu$_2$Si$_2$, the magnetic fluctuations of other heavy-fermion
systems across their metamagnetic transition have not yet been systematically investigated by neutron scattering. However, inelastic neutron
scattering at zero-field permitted to extract the antiferromagnetic linewidth for Ce$_{0.925}$La$_{0.075}$Ru$_2$Si$_2$ \cite{Knafo2009,Knafo2004},
CeNi$_2$Ge$_2$ \cite{Fak2000}, CeCu$_6$ \cite{RossatMignod1988}, and URu$_2$Si$_2$ \cite{Broholm1991}, whose value in these systems is rather close
to the value of $T_{\chi,max}$, as in CeRu$_2$Si$_2$ (see Table \ref{tablePM}). In CeIrIn$_5$ \cite{Willers2010} and USn$_3$ \cite{Loewenhaupt1990},
the linewidth measured on an averaged wavevector $\langle\mathbf{Q}\rangle$ is also similar to $T_{\chi,max}$ (see Table \ref{tablePM}), but it is
not known whether this linewidth results from an intersite signal or a local $Q$-independent signal. To test if $H_m$ and $T_{\chi,max}$ are
controlled by the onset of antiferromagnetic correlations in these systems, inelastic neutron scattering might be performed systematically to check
whether antiferromagnetic correlations disappear above $H_m$ (as in CeRu$_2$Si$_2$) or if a different behavior could be observed. As emphasized in
\cite{Scheerer2012}, the case of URu$_2$Si$_2$ above its hidden-order temperature $T_0=17.5$~K looks rather similar to the CeRu$_2$Si$_2$ case, since
the extrapolation of neutron data (measured up to 17~T) is compatible with a loss of antiferromagnetic correlations above 35~T \cite{Bourdarot2003}
(cf. also Sub-section \ref{ho}).

\begin{figure}[b]
\centering
\includegraphics[width=0.8\textwidth]{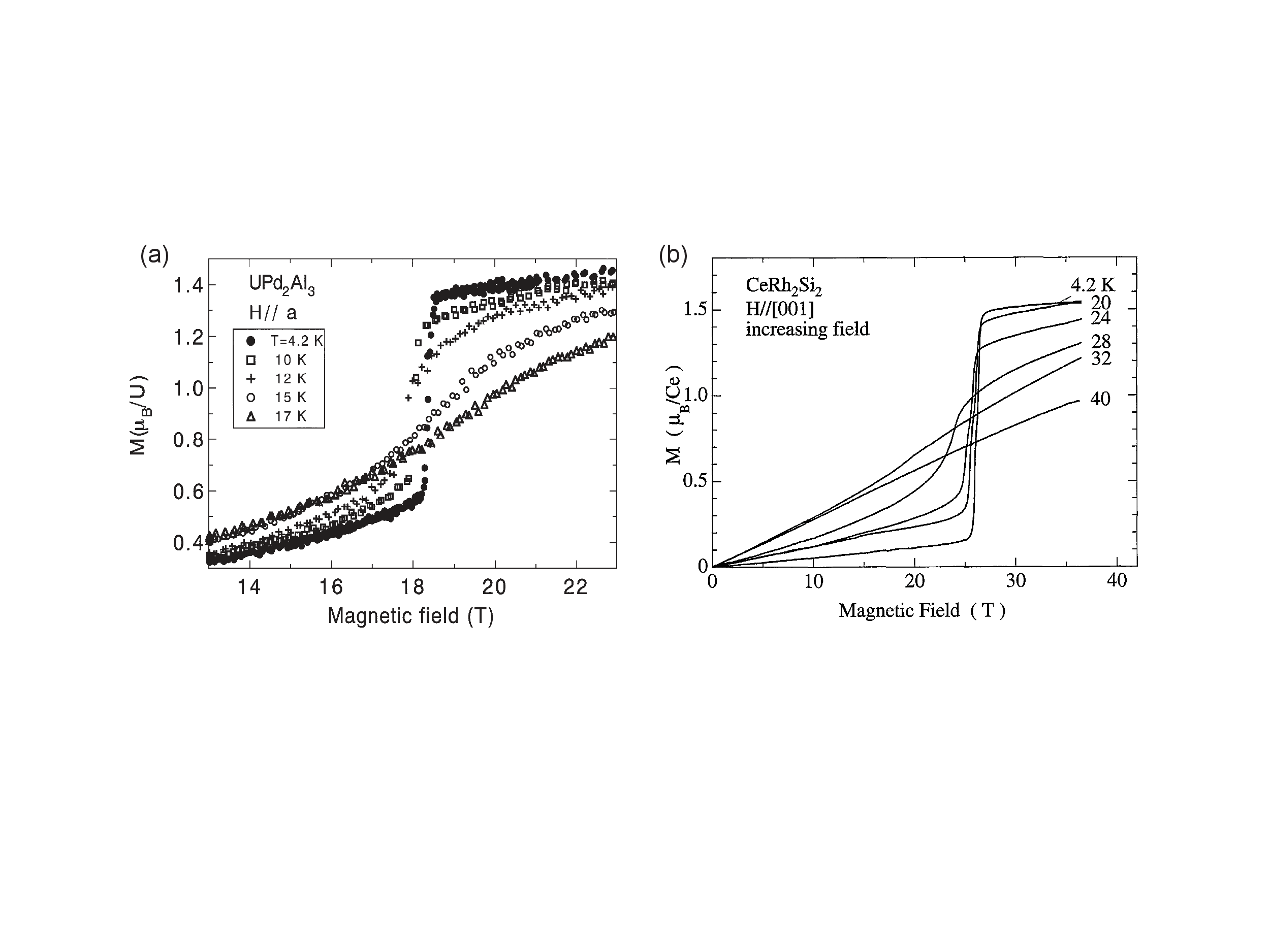}
\caption{High-field magnetization of (a) UPd$_2$Al$_3$ \cite{Sakon2002} and (b) CeRh$_2$Si$_2$ \cite{Settai1997}.}
\label{MHAF}
\end{figure}

\subsection{Heavy-fermion antiferromagnets}\label{af}

\begin{figure}[t]
\centering
\includegraphics[width=0.8\textwidth]{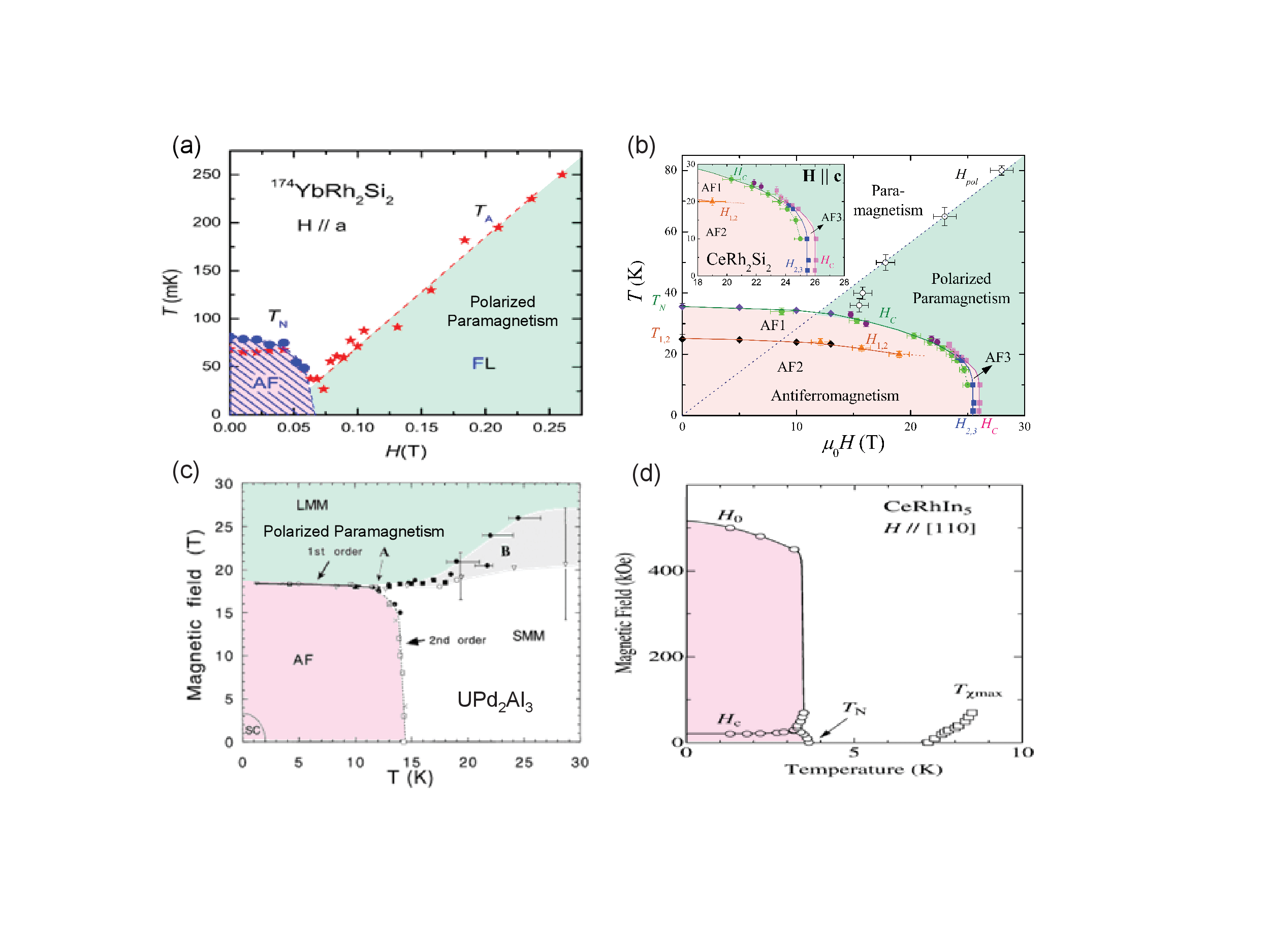}
\caption{Magnetic field-temperature phase diagrams of (a) YbRh$_2$Si$_2$ \cite{Knafo2010}, (b) CeRh$_2$Si$_2$ \cite{Knafo2010}, (c) UPd$_2$Al$_3$
\cite{Sakon2002}, with $\mathbf{H}\parallel\mathbf{c}$, and (d) CeRhIn$_5$ with $\mathbf{H}\perp\mathbf{c}$ \cite{Takeuchi2001}.} \label{diagallAF}
\end{figure}

\begin{table}[t]
\begin{center}
\caption{N\'{e}el temperature, critical magnetic field, and field direction for the metamagnetic transition in various heavy-fermion antiferromagnets.}
\begin{tabular}{lcccccc}
Material&$\:\:$&$T_N$ (K)&$\:\:$&$H_c$ (T)&$\mathbf{H}\parallel$&References\\
\hline
Ce$_{1-x}$La$_{x}$Ru$_2$Si$_2$ ($x=16\rightarrow70$~\%)&&$5.4\rightarrow3.2$&&$3.4\rightarrow1$&$\mathbf{c}$&\cite{Matsumoto2008,Matsumoto2010,Mignot1991}\\
CeRu$_2$(Si$_{1-x}$Ge$_{x}$)$_2$ ($x=10\rightarrow52$~\%)&&$6.6\rightarrow10.4$&&$3\rightarrow0.4$&$\mathbf{c}$&\cite{Matsumoto2011,Mignot1991}\\
Ce(Ru$_{0.92}$Rh$_{0.08}$)$_2$Si$_2$&&4&&3&$\mathbf{c}$&\cite{Aoki2012}\\
CeCu$_{2}$Ge$_2$&&4&&8&$[\overline{1} 1 0]$&\cite{Singh2011}\\
CeRh$_{2}$Si$_2$&&36&&26&$\mathbf{c}$&\cite{Knafo2010}\\
CeIn$_3$&&10&&61&(cubic)&\cite{Ebihara2004}\\
CeIn$_{2.75}$Sn$_{0.25}$&&6.4&&42&(cubic)&\cite{Silhanek2006}\\
CeRhIn$_5$&&3.8&&2/52&$[1 1 0]$&\cite{Takeuchi2001}\\
YbRh$_{2}$Si$_2$&&0.07&&0.06&$\mathbf{c}$&\cite{Knebel2006}\\
Yb$_3$Pt$_4$&&2.4&&1.85&$\perp\mathbf{c}$&\cite{Wu2011}\\
YbNiSi$_3$&&5.1&&8.3 (9)&$\mathbf{b}$ ($\perp\mathbf{b}$)&\cite{Avila2004,Grube2007}\\
U$_2$Zn$_{17}$&&9.7&&32&$[1 1 \overline{2} 0]$&\cite{Tateiwa2011}\\
UPd$_2$Al$_3$&&14&&18.4&$\mathbf{a}$&\cite{Sakon2002}\\
UPb$_3$&&32&&21&$\mathbf{a}$&\cite{Sugiyama2002}\\
UPt$_2$Si$_2$&&32&&45 (32)&$\mathbf{a}$ ($\mathbf{c}$)&\cite{Schulze2012}\\
\end{tabular}
\label{tableAF}
\end{center}
\end{table}

Due to a strong anisotropy, the field-induced transition of heavy-fermion antiferromagnets is often first-order, being accompanied by a sudden
step-like change in the magnetization. In UPd$_2$Al$_3$, antiferromagnetism sets up below $T_N=14$~K and a first-order transition is induced at
$\mu_0H_c=18.4$~T for $\mathbf{H}\parallel\mathbf{a}$ \cite{Sakon2002} [Figure \ref{MHAF} (a)]. As shown in Figure \ref{diagallAF} (c), the
antiferromagnetic-to-paramagnetic borderline of UPd$_2$Al$_3$ is of second-order at high-temperature and becomes first-order at low-temperature. Such
change from second- to first-order is also reported in other heavy-fermion antiferromagnets, as CeRh$_2$Si$_2$ \cite{Knafo2010} [Figure
\ref{diagallAF} (b)]. In this system, the N\'{e}el temperature $T_N=36$~K is one of the highest in the heavy-fermion family and a cascade of two
field-induced first-order phase transitions develops below 20~K, with presumably an antiferromagnetic canted phase stabilized between 25.5 and 26~T
\cite{Settai1997} [Figure \ref{MHAF} (b)]. Some exceptions, as YbNiSi$_3$, behave in a magnetic field as 'textbook' localized and almost-isotropic
antiferromagnets: YbNiSi$_3$ is characterized by a spin-flop transition at 1.7~T, when the field is applied along the easy magnetic axis
$\mathbf{b}$, and gets polarized above similar critical fields of 8.3 and 9.5~T, when the field is applied along and perpendicularly to $\mathbf{b}$,
respectively \cite{Avila2004,Grube2007}. For comparison with the cases of CeRh$_2$Si$_2$ and UPd$_2$Al$_3$, where the magnetic energy scales are high
[several tens of K or T, see Figure \ref{MHAF} (b-c)], Figure \ref{MHAF} (a) shows the magnetic field-temperature phase diagram of YbRh$_2$Si$_2$,
which has extremely small temperature and magnetic scales $T_N=70$~mK and $H_c=60$~mT, respectively \cite{Knebel2006}. It comes out that for
YbRh$_2$Si$_2$, CeRh$_2$Si$_2$, and UPd$_2$Al$_3$, despite very different -by a factor 500- energy scales, the critical field $H_c$ and the N\'{e}el
temperature $T_N$ scale rather well within a correspondence 1~T~$\leftrightarrow$~1~K. In these three systems, the characteristic temperature of the
high-field regime ($H>H_c$) increases almost linearly with $H$, being presumably related to the Zeeman energy.

\begin{figure}[t]
\centering
\includegraphics[width=0.6\textwidth]{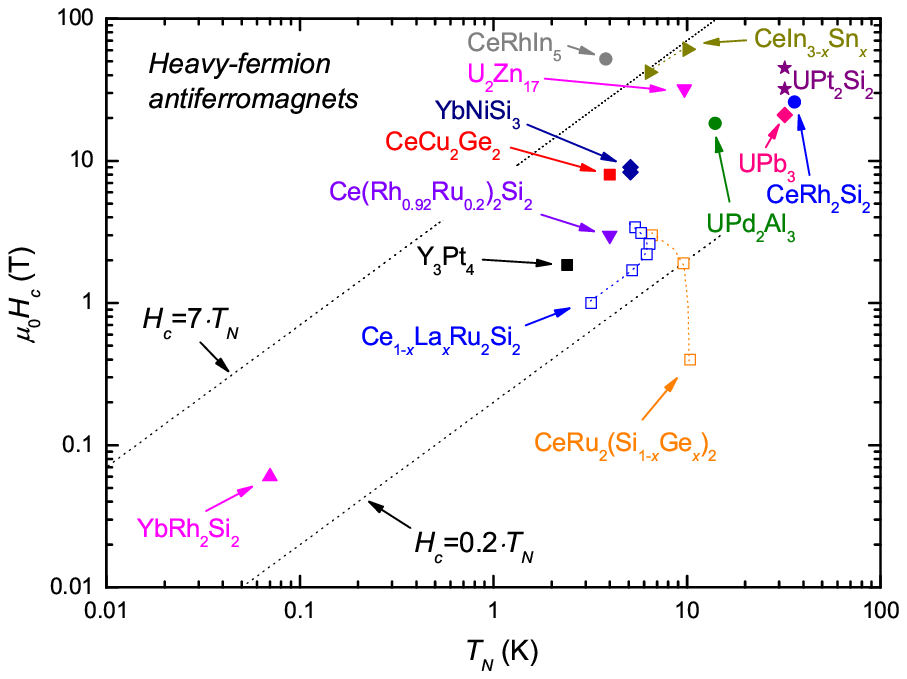}
\caption{Critical field as a function of the N\'{e}el temperature of various heavy-fermion antiferromagnets
\cite{Mignot1991,Aoki2012,Knafo2010,Knebel2006,Singh2011,Sakon2002,Tateiwa2011,Grube2007,Wu2011,Silhanek2006,Ebihara2004}.} \label{BTN}
\end{figure}

\begin{figure}[t]
\centering
\includegraphics[width=1\textwidth]{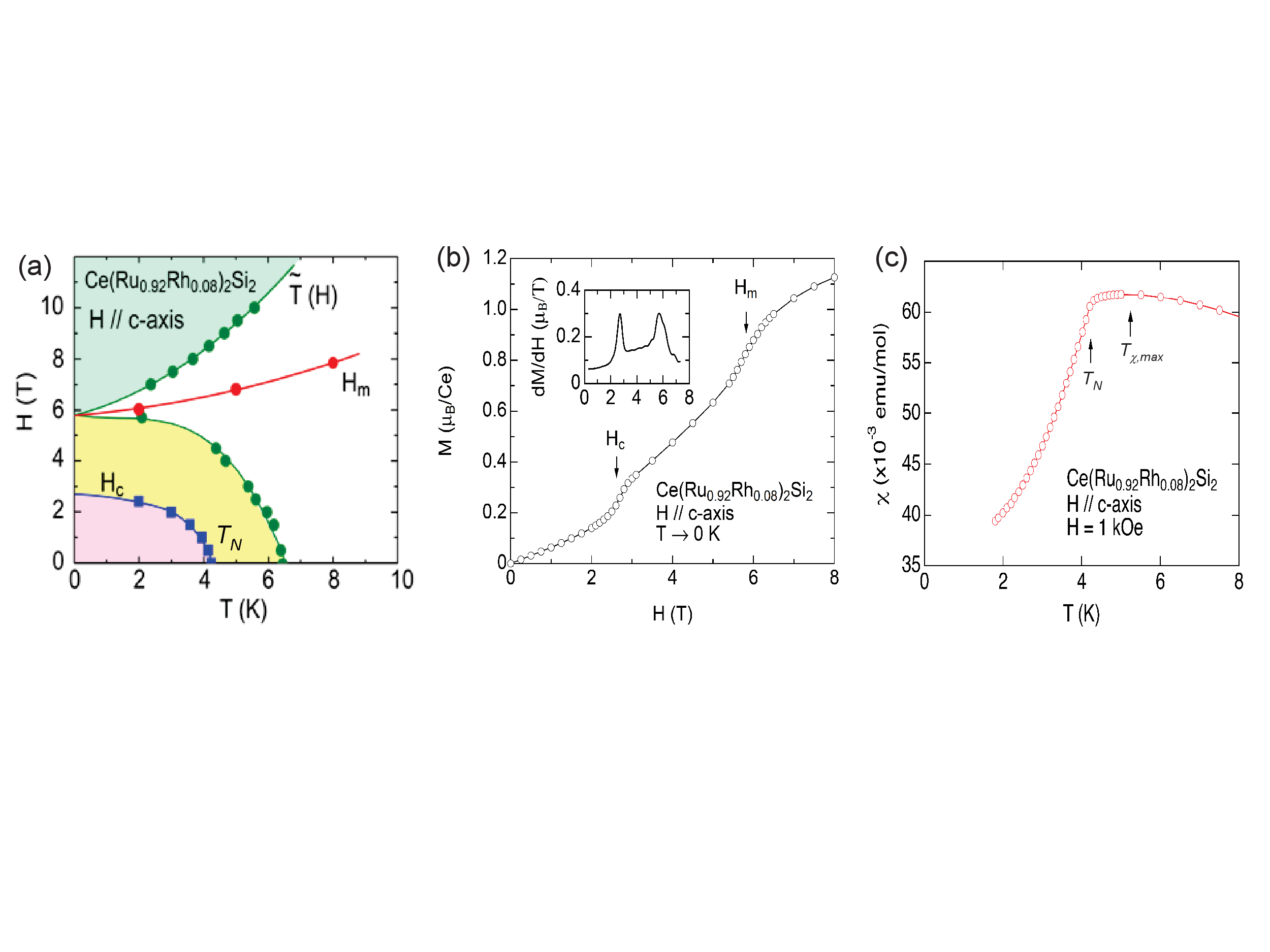}
\caption{(a) Magnetic field-temperature phase diagram, (b) magnetization versus field and (c) magnetic susceptibility versus temperature of
Ce(Rh$_{0.92}$Ru$_{0.08}$)$_2$Si$_2$ with $\mathbf{H}\parallel\mathbf{c}$ \cite{Aoki2012}.} \label{cerhrusi}
\end{figure}

Table \ref{tableAF} lists the values of $T_N$ and $H_c$ for different heavy-fermion antiferromagnets. $T_N(H=0)$ and $H_c(T\rightarrow0)$ are
believed to be controlled by the same exchange interactions, since the $H_c$ line is the continuation of the $T_N$ line in the $(H,T)$ plane. It is
thus natural to plot (in a log-log scale) $H_c$ versus $T_N$ in Figure \ref{BTN} instead of $H_c$ versus $T_{\chi,max}$ as in Figure \ref{BTchimax}
(cf. also \cite{Inoue2001,Fukuhara1996,Sugiyama2002,Onuki2004,Takeuchi2010}). Similarly to the $H_m$ versus $T_{\chi,max}$ plot made for
heavy-fermion paramagnets in Figure \ref{BTchimax}, the  $H_c$ versus $T_N$ plot shows that these two quantities are roughly related by a
correspondence 1~T~$\leftrightarrow1$~K, which indicates that they are mainly controlled by one energy scale. However, the scattering of the data
points in the $H_c$ versus $T_N$ plot (Figure \ref{BTN}) is more important than that in the $H_m$ versus $T_{\chi,max}$ plot (Figure \ref{BTchimax}):
most of the heavy-fermion antiferromagnets considered here are such that $0.2\cdot T_N<H_c<7\cdot T_N$, when $H_c$ and $T_N$ vary by almost three
orders of magnitude. This strong deviation from a unique scaling law illustrates the necessity to considerer several parameters to describe
heavy-fermion antiferromagnetism, instead of a single parameter if $H_c$ and $T_N$ would have fallen in a simple and unique law for all systems.
Additional parameters like multiple exchange paths, the dimensionality of these exchange couplings, the magnetic anisotropy, could explain the strong
scattering of the data points in Figure \ref{BTN}. For example, the case of CeRhIn$_5$ \cite{Takeuchi2001}, which is antiferromagnetically ordered
below $T_N=3.8$~K, deviates strongly from the 'mean' behavior of other heavy-fermion antiferromagnets: as shown in Figure  \ref{diagallAF} (c), its
phase diagram for $\mathbf{H}\parallel[1 1 0]$ is composed of two field-induced phase transitions at 2 and 52~T. A factor $\simeq14$~T/K instead of a
'mean' slope $\simeq1$~T/K relates the upper critical field to $T_N$. A low-dimensionality of the magnetic interactions proposed in
\cite{Takeuchi2001} to describe the field-induced increase of $T_{\chi,max}$ in CeRhIn$_5$, could also explain the strong deviation found here.

In heavy-fermion antiferromagnets, $T_{\chi,max}$ is not always equal to $T_N$, being sometimes higher. While the field-variation of $T_N$ has been
systematically studied in many heavy-fermion systems, that of $T_{\chi,max}$, when it differs from $T_N$, has generally not been characterized
carefully. The question is whether $T_{\chi,max}$ vanishes at the same field $H_c$ than $T_N$ or if it vanishes at a field $H_m$ higher than $H_c$.
Recently, a study of Ce(Ru$_{0.92}$Rh$_{0.08}$)$_2$Si$_2$ \cite{Aoki2012} permitted to show the decoupling between the field-instabilities at
$H_c=3$~T and $H_m=6$~T (cf. Figure \ref{cerhrusi}). In the magnetic-field-temperature plane, $H_c(T\rightarrow0)$ is clearly identified as the
prolongation of the N\'{e}el line $T_N(H)$, and $H_m$ is the prolongation of a crossover line $T_{\alpha,max}$ determined by thermal expansion
\cite{Aoki2012}, which reaches 6.5~K at zero-field and is related to $T_{\chi,max}\simeq5$~K [Figure \ref{cerhrusi} (c)]. In the parent compound
CeRu$_2$Si$_2$, $H_m=8$~T is the prolongation of a crossover line $T_{\alpha,max}\simeq T_{\chi,max}\simeq10$~K
\cite{Fisher1991,Paulsen1990,Flouquet2002}. The antiferromagnet Ce(Ru$_{0.92}$Rh$_{0.08}$)$_2$Si$_2$ has thus both the characteristics of the
above-mentioned heavy-fermion antiferromagnets, where $H_c$ scales with $T_N$, and that of the heavy-fermion paramagnets described in Sub-section
\ref{pm}, where $H_m$ scales with $T_{\chi,max}$. Microscopically, as well as static antiferromagnetic moments, which develop below $T_N$, are
quenched above $H_c$, we speculate that strong antiferromagnetic fluctuations, which would develop below $T_{\chi,max}$, would also collapse above
$H_m$. Inelastic neutron scattering experiments are now needed to test this hypothesis.

\subsection{Fermi surface modifications at a field-induced transition}\label{fsm}

\begin{figure}[t]
\centering
\includegraphics[width=0.9\textwidth]{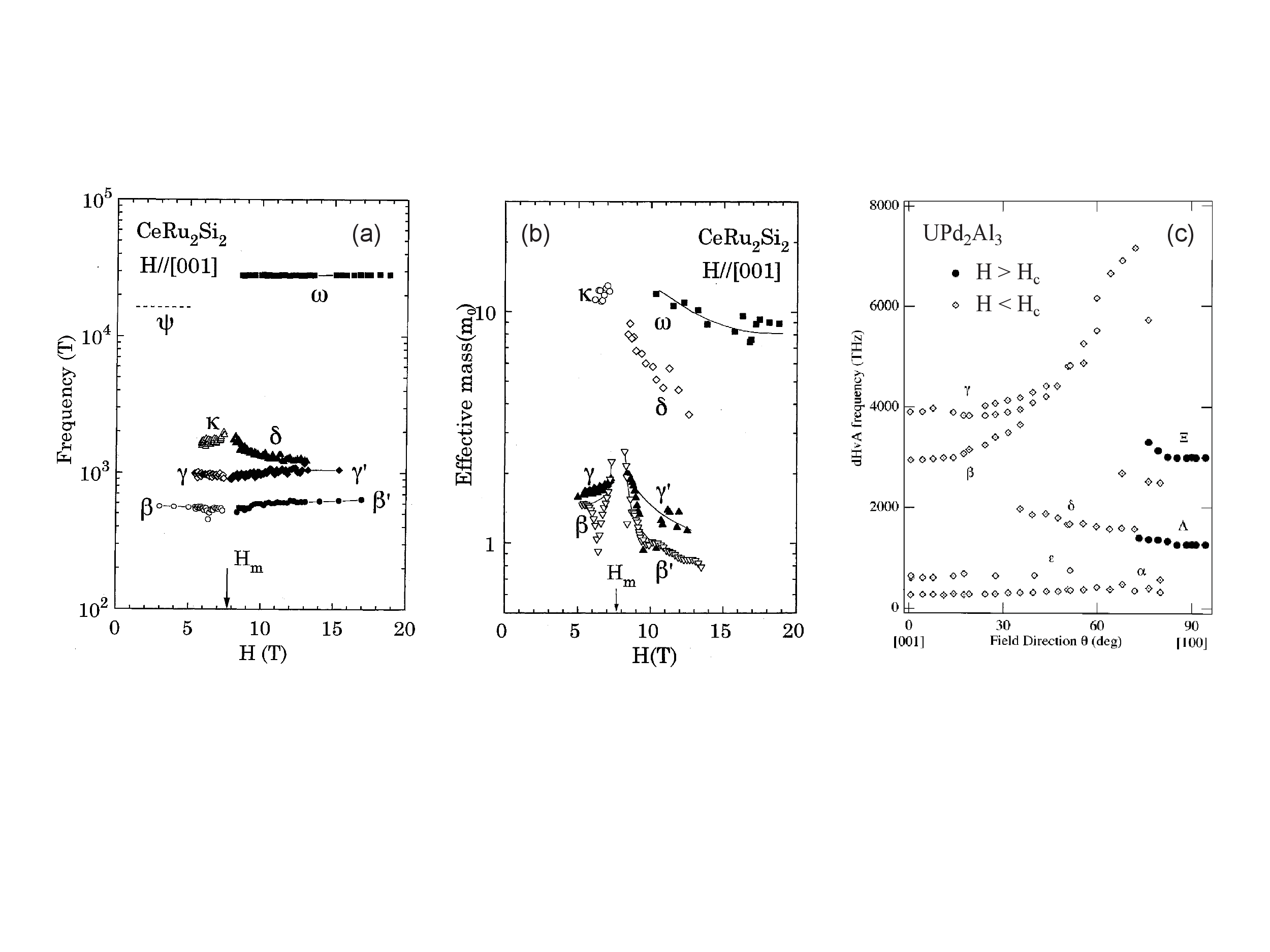}
\caption{Modifications of de Haas van Alphen (a) frequencies and (b) effective masses in the heavy-fermion paramagnet CeRu$_2$Si$_2$ in a magnetic
field applied along c \cite{Takashita1996} (c) Angular dependence of de Haas van Alphen frequencies at magnetic fields below and above the critical
field $H_c$ in the heavy-fermion antiferromagnet UPd$_2$Al$_3$ \cite{Terashima1997}.} \label{dHvA_Hc_Hm}
\end{figure}

In addition to modifications of the magnetic properties, a change in the Fermi surface has been observed by de Haas van Alphen oscillations at the
magnetic-field-induced transitions $H_m$ in the heavy-fermion paramagnet CeRu$_2$Si$_2$  \cite{Takashita1996} and $H_c$ in the heavy-fermion
antiferromagnet UPd$_2$Al$_3$ \cite{Terashima1997} (see Figure \ref{dHvA_Hc_Hm}). Controversially, the change of Fermi surface at $H_m$ in
CeRu$_2$Si$_2$ has been interpreted as the signature of a sudden localization of the $f$-electrons \cite{Takashita1996}, which is in contradiction
with the observation of local ($\mathbf{q}$-independent) Kondo magnetic fluctuations above $H_m$ by inelastic neutron scattering \cite{Raymond1999}.
Recent de Haas-van Alphen experiments \cite{Matsumoto2008} performed systematically on the whole family Ce$_{1-x}$La$_x$Ru$_2$Si$_2$ have also shown
a continuous evolution of the Fermi surface with doping in magnetic fields higher than $H_m$ and $H_c$. Hall effect and magnetoresistance
measurements on CeRu\(_2\)Si\(_2\) also support a continuous evolution of the Fermi surface through the field-driven magnetic transition
\cite{Daou2006}. Complementarily to the investigation of the Fermi surface, x-ray absorption spectroscopy is a pertinent tool to determine, via a
study of the $f$-ions valence, how the $f$ electrons become localized when a magnetic field is applied. This technique has been recently used for the
study of valence of the heavy-fermion materials CeCu$_2$Si$_2$ under pressure up to 8 GPa \cite{Rueff2011} as well as YbInCu$_4$ \cite{Matsuda2007}
and CeRu$_2$Si$_2$ \cite{Matsuda2012} under a pulsed magnetic field up to 40~T. Knowing that a high-enough magnetic field is expected to quench the
Kondo effect, leading to a localization of the $f$-electrons, x-ray absorption spectroscopy allows to determine whether the magnetic-field-induced
localization of the $f$-electrons suddenly occurs at a field-induced transition or progressively develops under magnetic field (with no marked
variation at $H_c$ or $H_m$). Concerning CeRu$_2$Si$_2$, it has been shown that the $f$-electron itineracy is progressively suppressed in a high
magnetic field applied along $\mathbf{c}$, with no Kondo breakdown at $H_m\simeq8$~T \cite{Matsuda2012}.

\subsection{The "Hidden-order" paramagnet URu\(_2\)Si\(_2\)}\label{ho}

\begin{figure}[t]
\centering
\includegraphics[width=0.9\textwidth]{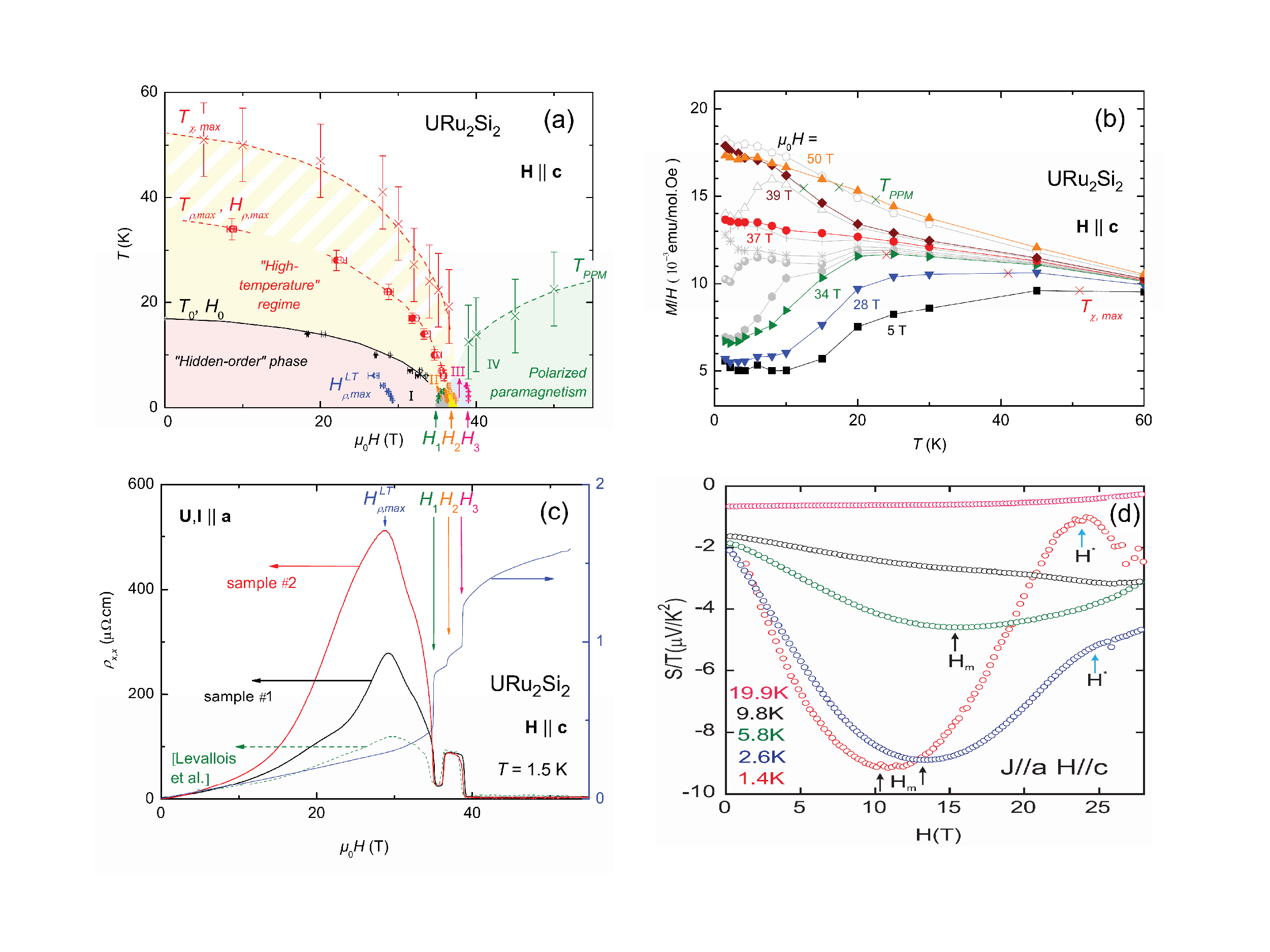}
\caption{ (a) Magnetic field-temperature phase diagram, (b) magnetization divided by the field versus temperature at different magnetic fields up to
50~T, and (c) magnetoresistance (of three single crystals with different RRR) and magnetization versus field at $T = 1.5$~K \cite{Scheerer2012}, and
(d) thermoelectric power divided by temperature versus magnetic field at different temperatures up to 27~T, for $\mathbf{H}\parallel\mathbf{c}$ in
URu$_2$Si$_2$ \cite{Malone2011}.} \label{urusi}
\end{figure}

The paramagnet URu$_2$Si$_2$ occupies a particular place in the heavy-fermion family. Despite a huge experimental and theoretical effort since more
than 20 years \cite{Mydosh2012}, the nature of the transition occurring at $T_0 = 17.5$~K and of its associated order parameter is still unknown.
Numerous models, as for example that based on antiferromagnetic hexadecapole ordering \cite{Kusunose2011}, have been proposed to described the
hidden-order phase below $T_0$. Enhanced magnetic fluctuations have been reported by inelastic neutron scattering at the wavevectors
$\mathbf{Q}_1=(1.4,0,0)$ and $\mathbf{Q}_0=(1,0,0)$ \cite{Broholm1991} and antiferromagnetic long-range ordering with the wavevector $\mathbf{Q}_0$
is stabilized above a pressure of 0.5~GPa \cite{Amitsuka2007}. A magnetic field along the easy magnetic axis $\mathbf{c}$ destabilizes the
hidden-order phase and replaces it, through a cascade of three first-order transitions, by a polarized paramagnetic regime above 39~T
\cite{Suslov2003} [Figure \ref{urusi} (a)]. The polarized magnetic moment reaches 1.5~$\mu_B$/U at 45~T and continues to increase significantly at
higher field \cite{Sugiyama1999a,Scheerer2012} [Figure \ref{urusi} (c)], showing that magnetic fluctuations are not fully quenched. Above $T_0$, the
high-field phase diagram looks rather similar to that of usual heavy-fermion paramagnets, as CeRu$_2$Si$_2$, since the field-induced transition
$H_m\simeq35-40$~T to a polarized regime corresponds to the prolongation in the ($T,H$) plane of the $T_{\chi,max}$ line. As shown in Figure
\ref{urusi} (b), $T_{\chi,max}$ decreases with $H$ and vanishes when the field-induced transition to a polarized regime occurs. The correspondence
1~K~$\leftrightarrow$~1~T between $T_{\chi,max}$ and $H_m$, as in many other heavy-fermion paramagnets (Figure \ref{BTchimax}), indicates that they
are controlled by a single energy scale (Table \ref{tablePM}). This scale might be related to antiferromagnet correlations, since $T_{\chi,max}=55$~K
is very close to the antiferromagnetic linewidth $\Gamma(\mathbf{Q}_0,T>T_0)\simeq50$~K probed by neutron scattering above $T_0$. Similarly to the
CeRu$_2$Si$_2$ case, an extrapolation of neutron data measured up to 17~T is also compatible with a loss above 35~T of antiferromagnetic correlations
\cite{Bourdarot2003}.

The singularity of the compensated metal URu$_2$Si$_2$ comes from the strong interplay between its magnetic properties and the properties of its
Fermi surface. When entering the "hidden-order" phase below $T_0$, the Fermi surface is suddenly modified, with a strong reduction of the carrier
number and an increase of the carrier mobility \cite{Scheerer2012,Dawson1989,Kasahara2007,Bel2004,Santander2009}. Below $T_0$, strong magnetic
fluctuations develop at the wavevector $\mathbf{Q}_0$, the linewidth of the magnetic fluctuations at $\mathbf{Q}_0$ and $\mathbf{Q}_1$ been also
significantly reduced \cite{Broholm1991}. No significant change of the Fermi surface has been seen in the pressure-induced antiferromagnetic state
\cite{Hassinger2010}, while successive modifications of the Fermi surface were observed when a magnetic field is applied along $\mathbf{c}$
\cite{Jo2008,Shishido2009,Altarawneh2011}, i.e., when substantial magnetic polarization is induced by the magnetic field. Figure \ref{urusi} (b)
shows that the transverse magnetoresistivity $\rho_{x,x}$ measured with $\mathbf{H}\parallel\mathbf{c}$ has a strongly-sample-dependent contribution
peaked at 30~T. This anomaly in the orbital contribution to $\rho_{x,x}$ is the signature of a progressive evolution from a high-mobility Fermi
surface below 30~T to a low-mobility Fermi surface above 30~T. A broad anomaly in the thermoelectric power at $\mu_0H^*=24$~T, as shown in Figure
\ref{urusi} (d), has also been attributed to a Lifshitz transition inside the hidden-order phase \cite{Malone2011}. In a high-magnetic field applied
along $\mathbf{c}$, the fact that $T_{\chi,max}/T_0$ is almost constant indicates that the destabilization of $T_{\chi,max}$ at 35-39~T drives the
destabilization of $T_0$ and that $T_{\chi,max}$ is a precursor of $T_0$. The unique cascade of three first-order transitions occurring in
URu$_2$Si$_2$ between 35 and 39~T is also a low-temperature consequence -of unknown origin- of the field-induced vanishing of $T_{\chi,max}$.

\section{Field- and spin-orientation-dependence of the effective mass}\label{effective_mass}

The dHvA effect has been successfully applied to numerous strongly correlated electron systems. In some heavy fermion compounds, dHvA measurements
have identified quasiparticles with masses of up to 100 times the bare electron mass~\cite{Aoki1992}, thus providing a direct evidence of a huge
quasiparticle density of states at the Fermi level, as is suggested by the strongly enhanced linear coefficient of the specific heat, \(\gamma\). In
many of the strongly correlated electron systems to which the dHvA technique has been applied~\cite{Onuki2003}, good agreement is found between
\(\gamma\) and the measured quasiparticle masses. There are notable exceptions, however,~\cite{Springford1991,Takashita1996} where the measured
quasiparticle masses add up to significantly less than the measured linear specific heat coefficient. Exotic theories of heavy fermion behavior
invoking neutral fermionic quasiparticles have been proposed to explain the "missing" mass in these systems~\cite{Kagan1992,Kagan1994,Tsvelik1992}.
Another theoretical approach is the possibility of a spin-split Fermi surface with an undetected heavy spin component. The latter possibility
received support from measurements by Harrison \emph{et al}. above the metamagnetic transition in CeB\(_6\)~\cite{Harrison1998}, which suggested that
quasiparticles of only a single spin orientation were contributing to the dHvA signal.

Experimentally, the magnetic field-dependence of the effective mass in heavy fermions was first observed in CeB\(_6\) in 1987. Joss \emph{et
al}.~\cite{Joss1987} performed low temperature dHvA measurements in CeB\(_6\) in fields up to 25 T. The authors observed a strong decrease of the
quasiparticle effective mass with magnetic field. However, the spin-orientation dependence of the effective mass was not observed. At about the same
time, Bredl~\cite{Bredl1987} reported that the low temperature specific heat coefficient of CeB\(_6\) also decreases with field up to 8 T.

From the experimental point of view, it is often difficult to follow the field-dependence of the spin-split effective masses. Small field intervals
required for the field-dependence are usually not sufficient to resolve the spin-splitting. Sometimes, the splitting can not be resolved directly
even over large field intervals. In this case, an attempt to extract the field-dependence of the effective mass by moving a small field window
through a large field range produces strange artificial results. A waveform analysis is then a more appropriate technique as will be discussed below.
Sometimes, however, the moving window method allows one to extract the field-dependence of the effective mass of both spin components.

Below, we will show two examples of heavy fermions compounds, CePd\(_2\)Si\(_2\) and CeCoIn\(_5\), where the spin-splitting of the Fermi surface can
be directly resolved in dHvA measurements and the effective masses of both spin components can be extracted and followed as a function of magnetic
field.

\subsection{CePd$_2$Si$_2$}

\begin{figure}[h]
\centering
\includegraphics[width=0.9\textwidth]{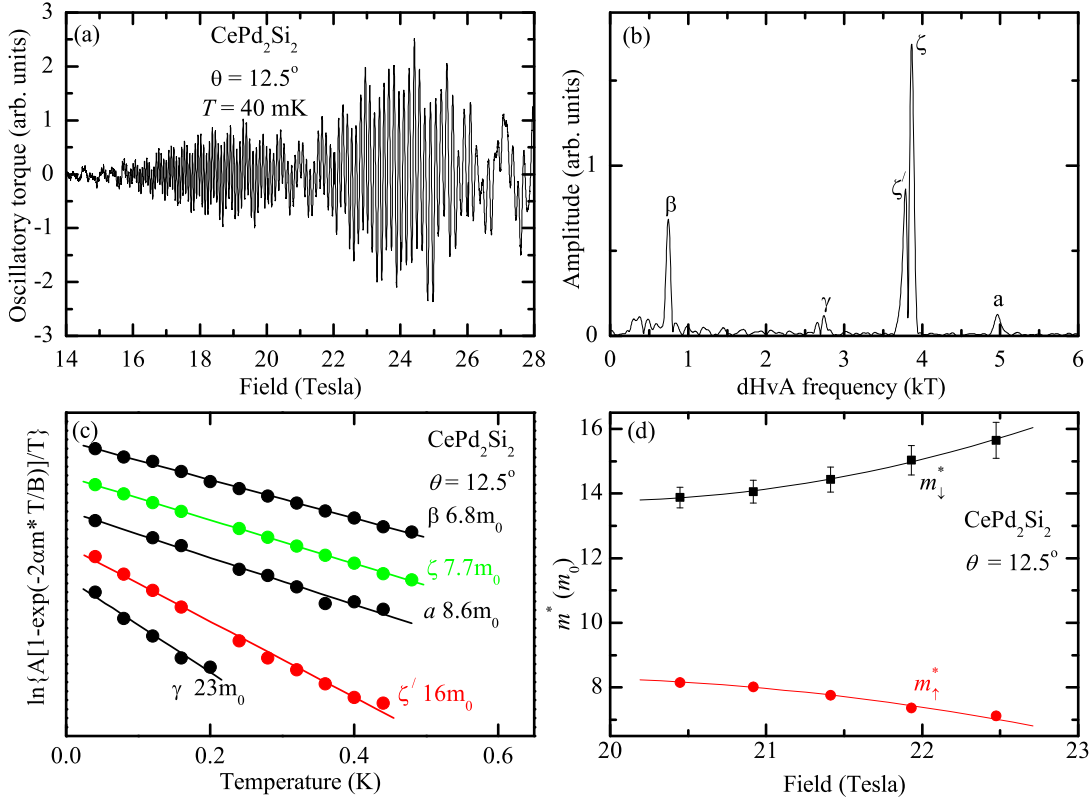}
\caption{dHvA oscillatory signal (a) and its Fourier spectrum (b) observed in CePd\(_2\)Si\(_2\) with magnetic field applied at 12.5\(^\circ\) from
[100] to [110] direction. Note the splitting of the \(\zeta\)-frequency. (c) Mass plot for the same orientation of the magnetic field. The slope of
the lines allows for the determination of the effective mass, which is found here to vary from 6.8 to 23\(\:m_0\). Note the two times difference
between the effective masses of the spin-split frequencies \(\zeta\) and \(\zeta '\). (d) Field dependence of both spin-up and spin-down cyclotron
effective mass. Lines are guide for the eye only illustrating the monotonic behavior of the effective masses of both spin-split bands above 20 T.}
\label{fig:CPS_12_5deg}
\end{figure}

Figures~\ref{fig:CPS_12_5deg} (a) and (b) show the dHvA signal and the corresponding Fourier spectrum in CePd\(_2\)Si\(_2\) with magnetic field
applied in the basal plane at 12.5\(^\circ\) from [100] direction. One of the fundamental peaks is split into two satellites \(\zeta\) and \(\zeta
'\) with frequencies 3.87 and 3.79 kT respectively. Such a small splitting can be resolved only when analyzing the data over a large inverse field
range. The two close satellites originate from the spin-splitting of the Fermi surface into up and down spin bands.

It is well known that in ferromagnetic metals, the band structure calculations usually predict very different majority and minority spin Fermi
surfaces. The effective masses corresponding to the up and down spin orientations are, therefore,  also very different. In other materials, however,
for a spin-splitting, two sheets of the Fermi surface with opposite orientation of the spins usually have similar shapes and topologies, and thus
close values of the effective mass. In the case of CePd\(_2\)Si\(_2\), however, the effective mass corresponding to the \(\zeta '\)-peak,
16\(\:m_0\), is more than two times higher than that of the other satellite. This can be seen unambiguously in figure~\ref{fig:CPS_12_5deg} (c)
showing the mass plot.

Such a remarkable difference would be surprising in other metals, but is rather expected in heavy fermion compounds, where different masses are
predicted theoretically for up- and down-spin bands. A similar behavior has been observed in PrPb$_3$,~\cite{Endo2003} where the effective masses of
up and down spin oscillations were found to be very different, up to a factor of two, although the frequencies were also very close to each other.
Such a big difference in the effective masses might shed some light on the puzzling situation in some other \(f\)-electron compounds, e.g.
CeB\(_6\),~\cite{Harrison1998,Goodrich1999} where the oscillations from only one spin state were observed. Indeed, this would be the case if the
other spin channel had too large effective mass to be detected.

Direct observation of the splitting of the \(\zeta\)-frequency [figure~\ref{fig:CPS_12_5deg} (b)] provides an opportunity to trace the
field-dependence of the effective mass of both spin components. The field-dependence of both up- and down-spin effective mass of the \(\zeta\)-orbit
is shown in figure~\ref{fig:CPS_12_5deg} (d). The analysis was limited to fields above 20 T, where the splitting of the \(\zeta\)-orbit could be
resolved unambiguously. The effective mass of the spin down band increases continuously with increasing field and that of the spin down band shows a
monotonic decrease. This behavior is in good qualitative agreement with modern theoretical
models~\cite{Spalek2006,Spalek2006a,Onari2008,ViolaKusminskiy2008}. An extension of these measurements to higher fields is required to check whether
the effective masses of both spin components still depend on field monotonically above 28 T.

\subsection{CeCoIn$_5$}

Previous moderate field dHvA studies of CeCoIn\(_5\)~\cite{Settai2001,Hall2001} have revealed essentially all the Fermi surface sheets predicted by
theoretical band structure calculations performed for itinerant \(f\)-electron. The thermodynamics is dominated by two large, quasi-2D sheets,
\(\alpha\) and \(\beta\). Strongly enhanced effective masses up to 87\(\:m_0\) were found to be field-dependent.

More recently, the quasiparticle masses of the thermodynamically important \(\alpha\) and \(\beta\) sheets of the Fermi surface were studied in more
details by McCollam \emph{et al}.~\cite{McCollam2005}. The authors analyzed temperature dependence of the dHvA oscillations over the field range of
13 to 15 T and down to very low temperature of 6 mK. They demonstrated that the conventional Lifshitz-Kosevich formula fails to provide a
satisfactory fit of the data. However, when spin-dependent effective masses were taken into account, the modified ``spin-dependent''
Lifshitz-Kosevich theory provided an excellent fit. It should be emphasized, however, that McCollam \emph{et al}. have not observed the
spin-splitting of the oscillatory signal directly. The spin-split effective masses were extracted from the analysis of the temperature dependence of
the oscillatory signal of a single frequency.

\begin{figure}[h]
\centering
\includegraphics[width=0.9\textwidth]{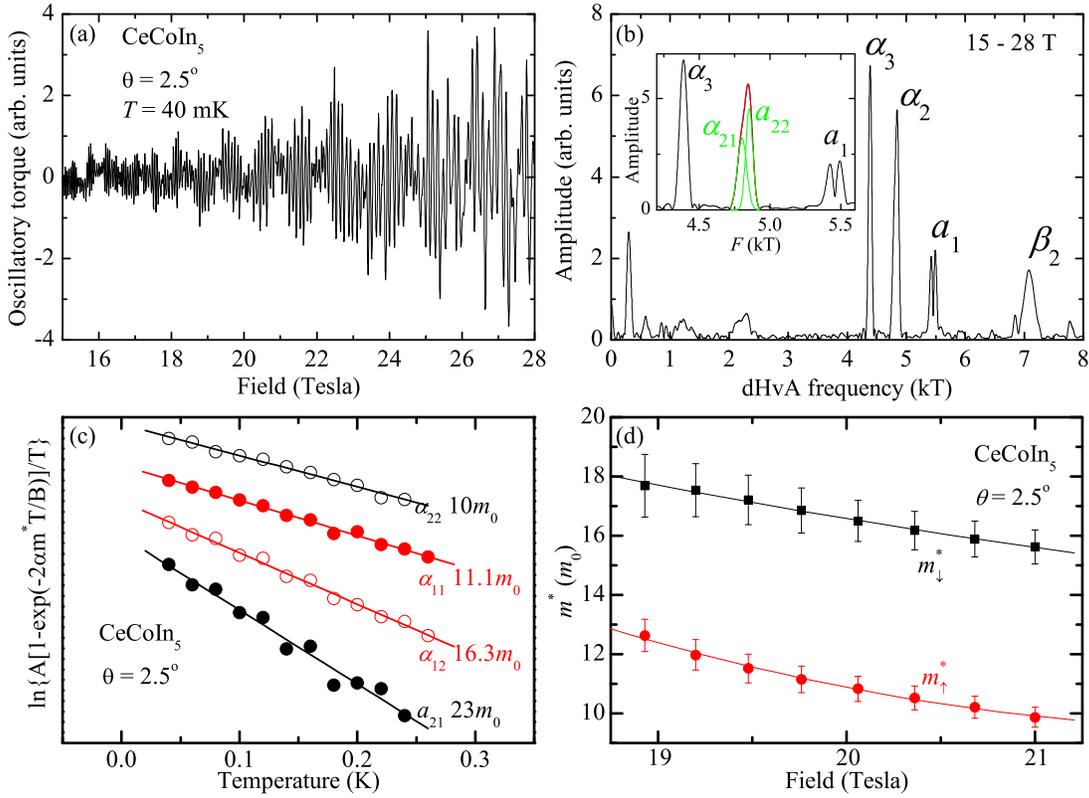}
\caption{dHvA oscillatory signal (a) and its Fourier spectrum from 15 to 28 T (b) observed in CeCoIn\(_5\) with magnetic field applied at
2.5\(^\circ\) from [100] to [100] direction. The spin-splitting of the \(\alpha_1\) and \(\beta_2\) frequencies can be seen clearly. The inset zooms
on the frequencies originating from the \(\alpha\) sheet of the Fermi surface. The characteristic shape and broadening of the \(\alpha_2\) peak
implies that it is composed of two close satellites. The satellites (green lines) are indeed revealed by fitting the peak by a two-peak Gaussian
function (red line). (c) Mass plot for the spin-split frequencies \(\alpha_1\) (red) and \(\alpha_2\) (black) for the same orientation of the
magnetic field. Open and close symbols correspond to the different spin components. (d) Effective masses of both spin components of the
\(\alpha_1\)-orbit as a function of applied field. Lines are guide for the eye.} \label{fig:CCI_2.5deg}
\end{figure}

Our dHvA measurements in CeCoIn\(_5\) were performed at field up to 28 T and temperature down to 40 mK. While the base temperature of our experiment
was not nearly as low as that of McCollam \emph{et al}.~\cite{McCollam2005}, a larger inverse field range of our measurements resulted in a much
better dHvA frequency resolution. The oscillatory signal obtained with magnetic field applied at 2.5\(^\circ\) from \(c\) to \(a\)-axis at \(T = 40\)
mK is shown in figure~\ref{fig:CCI_2.5deg} (a). The corresponding Fourier transform for a field range from 15 to 28 T is shown in
figure~\ref{fig:CCI_2.5deg} (b). The splitting of the \(\alpha_1\)-frequency is quite clear without any further analysis. The values of the split
frequencies, 5.42 and 5.49 kT, are very close to each other and differ by only about 1\%. It is not surprising, therefore, that the splitting could
not be resolved in previous measurements over a much smaller inverse field range.

At a first glance, the \(\alpha_2\) peak does not appear to be split. Its broadening and characteristic asymmetric shape, however, indicate that it
consists of two very close peaks. This can be seen in the inset of figure~\ref{fig:CCI_2.5deg} (b) showing a zoom of the frequencies originating from
the \(\alpha\) sheet of the Fermi surface. Indeed, an attempt to fit the \(\alpha_2\) peak with a double-peak Gaussian function (the red line)
produces an excellent result and reveals two close satellites (the green lines). The frequencies of the satellites, 4.80 and 4.85 kT, also differ by
about 1\%.

Figure~\ref{fig:CCI_2.5deg} (c) shows the mass plot of the split frequencies \(\alpha_1\) and \(\alpha_2\) from figure~\ref{fig:CCI_2.5deg} (b). The
spin-dependent effective masses of the \(\alpha_1\) orbit, \(11.1 \pm 0.3\:m_0\) and \(16.3 \pm 0.4\:m_0\), differ by about 50\%. However, the
difference between the effective masses of the \(\alpha_2\) satellites, \(\alpha_{21}\) and \(\alpha_{22}\), \(23.4 \pm 1.3\:m_0\) and \(9.9 \pm
0.3\:m_0\) is more than a factor of two. This ratio is similar to that observed in CePd\(_2\)Si\(_2\) as discussed above. The ratio of the two masses
is also similar to that found by McCollam \emph{et al}.~\cite{McCollam2005} for the \(\alpha_3\) orbit at 13--15 T.

The field-dependence of the effective masses corresponding to both up- and down-spin bands of the \(\alpha_1\)-orbit is shown in
figure~\ref{fig:CCI_2.5deg} (d). Contrary to CePd\(_2\)Si\(_2\), here both the up-spin and down-spin effective masses vary monotonically with field.
However, surprisingly, both masses are found to decrease with increasing field. The effective mass of the spin-up band decreases faster, changing by
21\% from 19 to 21 T. The effective mass of the spin-down-band varies by 12\% over the same field range.

The decrease of both spin components masses with magnetic field observed here agrees with the results of McCollam \emph{et al}.~\cite{McCollam2005}
who reported the same behavior, although their results were obtained at much lower fields. At a first glance, this behavior seems to be at odds with
the existing modern theories, as all of them predict an increase of one of the spin-dependent masses and a decrease of the other one. However, very
recently it was shown~\cite{Howczaka2011} that in ferromagnetic state of heavy fermion compounds, the application of sufficiently strong magnetic
field leads to a full polarization of the system. Then both spin-up and spin-down masses decrease with increasing magnetic field. This might explain
the unusual behavior of the effective masses observed here, even though CeCoIn\(_5\) is non-magnetic at zero magnetic field. However, neutron
diffraction measurements in magnetic fields above 20~T have not yet been performed on CeCoIn\(_5\).

\section{Non-centrosymmetric compounds}\label{Non-centrosymmetric}

\begin{figure}[h]
\centering
\includegraphics[width=0.4\textwidth]{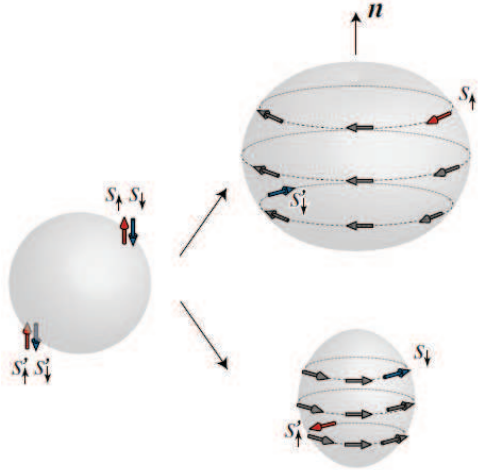}
\caption{Schematic illustration of the Fermi surface splitting in non-centrosymmetric materials. The original Fermi surface (in the left) is split
into two surfaces characterized by different chirality (in the right).} \label{fig:FS splitting}
\end{figure}

Non-centrosymmetric materials is a subclass of heavy fermion compounds. The absence of inversion symmetry in the crystal lattice of a metal brings
about a strong spin-orbit coupling, which in turn leads to the splitting of the electronic energy bands. The Fermi surface of such a metal is,
therefore, also split into two surfaces characterized by different chirality as schematically shown in figure~\ref{fig:FS splitting}. This naturally
results in the appearance of two distinct frequencies in the spectra of dHvA oscillations. It was demonstrated theoretically~\cite{Mineev2005} that
the analysis of the oscillatory spectra in such materials provides a direct measure of the strength of the spin-orbit coupling. It is thus very
important to obtain detailed and precise dHvA frequencies and their angular dependence in materials without inversion center.

Here, we illustrate the experimental observation of the Fermi surface splitting by dHvA measurements in non-centrosymmetric CeCoGe\(_3\) performed in
field up to 28 T. The dHvA effect in CeCoGe\(_3\) was previously investigated in magnet fields up to 17 T~\cite{Thamizhavel2006}. Four fundamental
frequencies were identified, all of them split due to spin-orbit interaction. The frequencies themselves and their angular dependencies are in a
rather good agreement with the results of theoretical band structure calculations performed for LaCoGe\(_3\), implying localized \(f\)-electrons. As
compared to other non-centrosymmetric compounds, the splitting of dHvA frequencies in CeCoGe\(_3\) is relatively small, indicating a moderate
spin-orbit coupling of about 100 K. The highest effective mass observed in CeCoGe\(_3\) is 12 bare electron masses corresponding to the
\(\beta\)-branch. Finally, only one frequency originating from the \(\alpha\)-branch representing the biggest Fermi surface was initially observed in
CeCoGe$_3$.

\begin{figure}[h]
\centering
\includegraphics[width=0.95\textwidth]{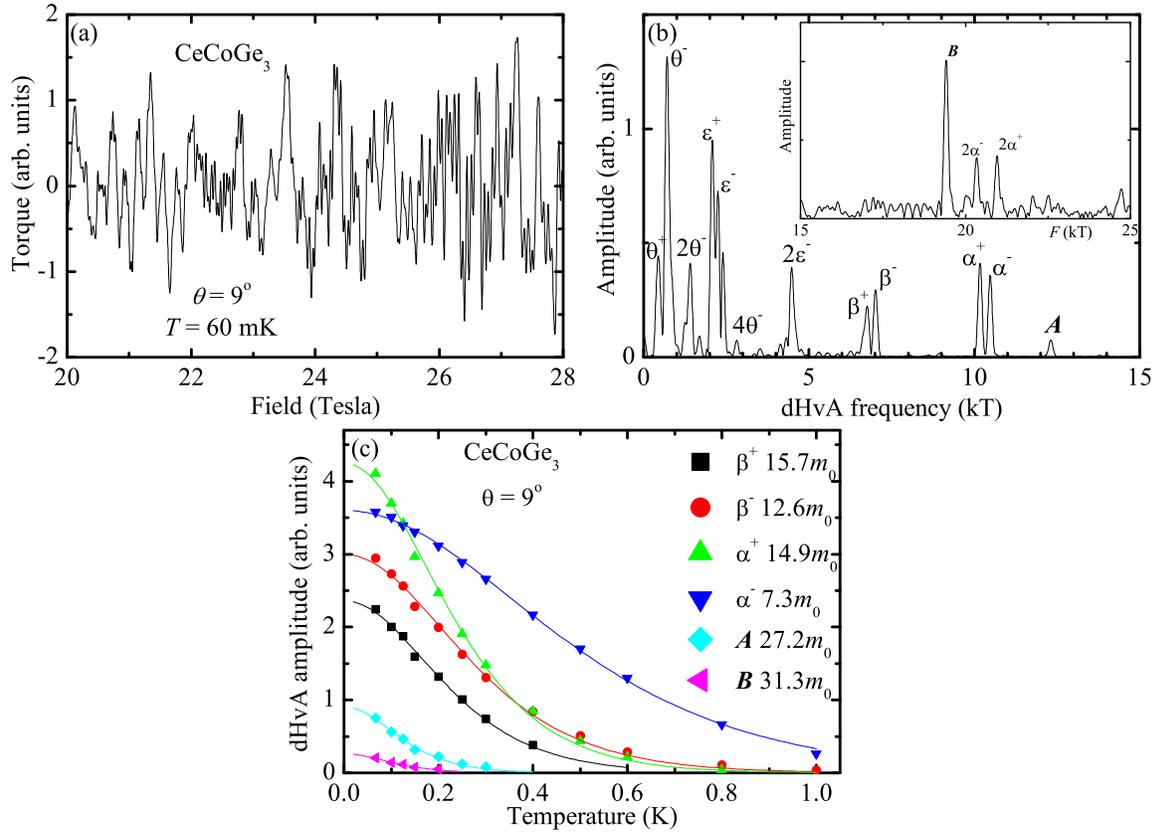}
\caption{dHvA oscillatory signal (a) and its Fourier spectrum (b) observed in CeCoGe\(_3\) with magnetic field (20---28 T) applied at 9\(^\circ\)
from [100] to [110] at \(T = 60\) mK. The frequencies denoted \(\theta\), \(\epsilon\), \(\beta\) and \(\alpha\) were observed in the previous lower
field measurements (with the exception of \(\alpha^+\)-branch). The frequencies denoted $A$ and $B$ are observed only at high field and were not
detected in the previous measurements. (c) Temperature dependence of the dHvA amplitude is shown for \(\alpha\) and \(\beta\)-branches as well as for
the new frequencies \(A\) and \(B\). Lines are the fits to the temperature dependent part of the Lifshitz-Kosevich formula. The effective masses,
\(m^\ast\), obtained from the fits are also shown.} \label{fig:CCG_dHvA&mass_9deg}
\end{figure}

The dHvA oscillations observed in CeCoGe\(_3\) between 20 and 28 T are shown in figure~\ref{fig:CCG_dHvA&mass_9deg} (a). All the fundamental
frequencies previously observed in CeCoGe$_3$ are still present at high field [figure~\ref{fig:CCG_dHvA&mass_9deg} (b)]. In addition, both components
of the \(\alpha\)-branch are now clearly resolved. With magnetic field applied at 9\(^\circ\) from the crystallographic \(c\)-axis, the two
frequencies are close to each other, \(F_{\alpha^+} = 10.17\) kT and \(F_{\alpha^-} = 10.47\) kT. The most significant and surprising result,
however, is the presence of the new fundamental frequencies, \(A\) and \(B\), in the Fourier spectrum of CeCoGe\(_3\). The new frequencies, \(F_A =
12.31\) kT and \(F_B = 19.42\) kT are considerably higher than the highest frequency, \(F_\alpha\), reported for lower fields.

Figure~\ref{fig:CCG_dHvA&mass_9deg} (c) shows the temperature dependence of the dHvA amplitudes of the new frequencies \(A\) and \(B\) as well as
\(\alpha\) and \(\beta\)-branches in CeCoGe\(_3\) for magnetic field applied at 9\(^\circ\) from the crystallographic \(c\)-axis. These data allow
one to determine effective masses by fitting the experimental points to the temperature-dependent part of the Lifshitz-Kosevich formula. The best
fits to the formula along with the extracted effective masses are also shown in figure~\ref{fig:CCG_dHvA&mass_9deg} (c). It was previously
reported~\cite{Thamizhavel2006} that \(\alpha\) and \(\beta\)-branches possess the highest effective masses of 8 and 12 bare electron masses
respectively. These values are very close to the current results if only the \(\alpha^-\) frequency is considered and taking into account that only a
single frequency from the \(\alpha\)-branch was initially observed in previous measurements. Interestingly, the effective masses of the two
frequencies originating from the \(\alpha\)-branch differ by a factor of more than two, being 14.9 and 7.3 bare electron masses for \(\alpha^+\) and
\(\alpha^-\) frequencies respectively. This is in contrast with all the other branches where the effective masses of the two components are quite
close to each other. The most surprising result, however, is that the effective masses of the new frequencies are strongly enhanced being \(27\:m_0\)
and \(31\:m_0\) for \(A\) and \(B\) respectively. These values by far exceed the masses of the other previously observed frequencies.

For the moment, it is not clear where the two new high dHvA frequencies observed in CeCoGe$_3$ originate from. Neither of them was detected in the
previous lower-field measurements or revealed by theoretical band structure calculations. The frequencies correspond to strongly enhanced effective
masses implying that they represent thermodynamically important parts of the Fermi surface. While it is not certain if the frequencies appear in high
magnetic field only or are simply experimentally undetectable at lower field, they certainly do not originate from magnetic breakdown. They are,
therefore, likely to be intrinsic and are possibly due to a field-induced modification of the Fermi surface, especially as similar new frequencies
were also detected in the non-4\(f\) analog LaCoGe\(_3\)~\cite{Sheikin2011}. It would be interesting to see if new frequencies emerge in other
non-centrosymmetric compounds at high field.

\section{U-based ferromagnetic superconductors}\label{FM}

Ferromagnetism (FM) and superconductivity (SC) are usually antagonistic, because the strong internal field due to ferromagnetism generally easily
destroys superconductivity. In early 1980s' extensive studies had been done for the exceptional cases of several materials, such as Chevrel phase
compounds (REMo$_6$Se$_8$, RE: rare earth) and ErRh$_4$B$_4$, where the magnetic moment is relatively large and SC phase is expelled when the FM
state is established.~\cite{Fischer1990} In this case, the Curie temperature ($T_{\rm Curie}$) is lower than the superconducting critical temperature
($T_{\rm sc}$). In the special cases where the SC coherence length $\xi$ is larger than the size of magnetic domain $d$, SC can coexist with FM in a
narrow temperature range. At lower temperatures, SC is destroyed, which illustrates that SC and FM compete with each other.

The microscopic coexistence of FM and SC was theoretically predicted for the well-known weak ferromagnet ZrZn$_2$~\cite{Fay1980}, where a
spin-triplet state with equal spin pairing was expected to be formed near the ferromagnetic quantum critical point. However, there are no
experimental evidences for SC in this material, although extrinsic SC due to Zr-alloys on the surface was reported~\cite{Yelland2005}.

The first discovery of coexistence of FM and SC was reported in UGe$_2$~\cite{Saxena2000}, where SC is observed only in the FM phase just below the
critical pressure at which FM is suppressed. URhGe is the first ferromagnet which shows SC already at ambient pressure~\cite{Aoki2001}. $T_{\rm
Curie}$ ($=9.5\,{\rm K}$) is much higher than $T_{\rm sc}$ ($=0.25\,{\rm K}$), indicating that the SC phase exists in the FM phase. Coexistence of SC
and FM is also found in UCoGe,~\cite{Huy2007} which has the same crystal structure than URhGe. The weak ferromagnet UIr also shows SC just below the
critical pressure of ferromagnetism~\cite{Akazawa2004}. The ($T,P$) phase diagram with multiple FM phases is more complicated, and bulk SC
properties, for example in the specific heat, have not been confirmed yet. Here we do not describe the details of UIr further.

\begin{figure}[bbv]
\begin{center}
\includegraphics[width=1 \hsize,clip]{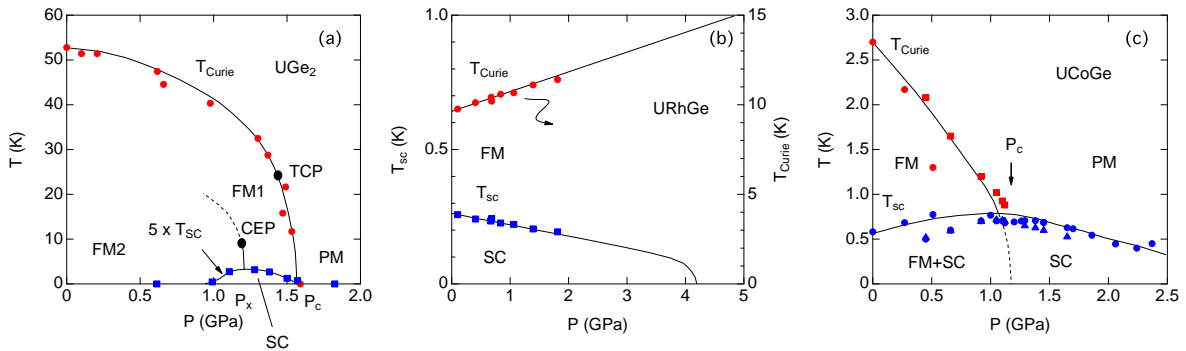}
\end{center}
\caption{($T,P$) phase diagram of (a) UGe$_2$, (b) URhGe and (c) UCoGe.~\cite{Saxena2000,Taufour2011,Huxley2001,Hardy2005,Miyake2009,Hassinger2008}}
\label{fig:TP_phase}
\end{figure}

All the known materials where FM coexists microscopically with SC are uranium compounds with substantially reduced magnetic moments compared to the
free ion values. The 5$f$-electrons in these systems have in general an intermediate nature between itinerant 3$d$-electrons and localized
4$f$-electrons. The magnetic moment of uranium compounds varies from nearly free ion values (localized case) to the very tiny values (itinerant
case). Furthermore, a strong spin-orbit interaction also plays an important role for the physical properties. The ferromagnetic superconductors
mentioned above have relatively small magnetic moments, thus 5$f$-electrons in these compounds are thought to be itinerant at first approximation.

\subsection{UGe$_2$ and the ferromagnetic quantum critical end point}

Figure~\ref{fig:TP_phase}(a) shows the ($T,P$) phase diagram of UGe$_2$~\cite{Saxena2000,Taufour2011,Huxley2001}. The large $T_{\rm Curie}$ at
ambient pressure in UGe$_2$ is suppressed by applying pressure. At the critical pressure $P_{\rm c}\sim 1.5\,{\rm GPa}$, FM completely collapses and
a paramagnetic (PM) ground state is realized. SC is observed only in the FM regime just below $P_{\rm c}$. The maximum of $T_{\rm sc}$ is found at
$P_{\rm x}\sim 1.2\,{\rm GPa}$, where $T_{\rm x}$ collapses. FM below $T_{\rm Curie}$ consists of two phases FM1 and FM2, which are separated by
$T_{\rm x}$. The phase FM1 between $T_{\rm Curie}$ and $T_{\rm x}$ is weakly polarized ($M_0 \sim 1\,\mu_{\rm B}$), while the phase FM2 below $T_{\rm
x}$ is strongly polarized ($M_0\sim 1.5\,\mu_{\rm B}$). At low pressure the boundary between FM1 and FM2 is a crossover, and at the pressure $P_{\rm
CEP}$ (just below $P_{\rm x}$) it changes into a first-order transition, at a point called critical end point (CEP). $T_{\rm Curie}$ also changes
into a first-order transition at the tri-critical point (TCP) at $P_{\rm TCP}$, which is just below $P_{\rm c}$.

\begin{figure}[bbv]
\begin{center}
\includegraphics[width=1 \hsize,clip]{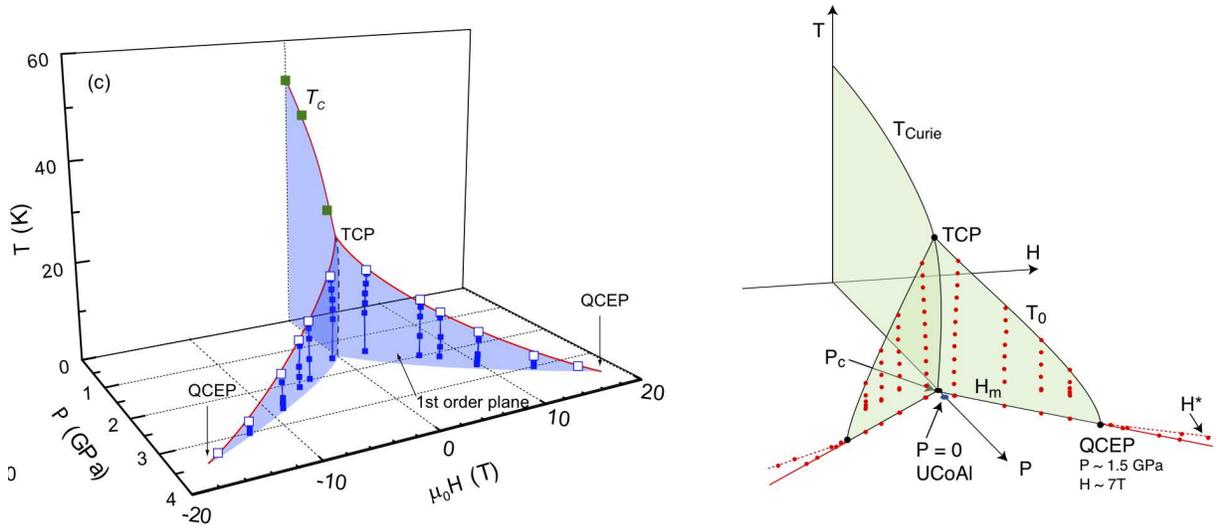}
\end{center}
\caption{($T,P,H$) phase diagram of (a) UGe$_2$~\cite{Taufour2010,Kotegawa2011} and (b) UCoAl~\cite{Aoki2011a}.}
\label{fig:QCEP}
\end{figure}

When the magnetic field is applied along the $a$-axis (easy-magnetization axis) at $P > P_{\rm c}$, the phase FM1 is recovered from the low-field PM
phase through a sharp first-order metamagnetic transition. At $P_{\rm x} < P < P_{\rm c}$, a step-like first-order metamagnetic transition from FM1
to FM2 is also observed. The ($T,P,H$) phase diagram of UGe$_2$ is shown in Fig.~\ref{fig:QCEP}(a). The critical temperature decreases with field
from the TCP defined at zero field and terminates at the quantum critical end point (QCEP, at $P_{\rm QCEP}\sim 3.5\,{\rm GPa}$, $H_{\rm QCEP}\sim
18\,{\rm T}$). By crossing the first-order plane (between the TCP and the QCEP, for $T$ below the critical temperature), we can observe a sharp
metamagnetic transition. A similar ``wing''-shaped phase diagram is also realized between FM1 and FM2, however the critical field at which the
first-order transition terminates is probably much higher.

de Haas-van Alphen (dHvA) experiments clearly demonstrate that the Fermi surfaces in FM1, FM2 and in the PM phase are quite
different~\cite{Terashima2001,Settai2002}. For example, for $H\parallel b$, dHvA branches present in FM2 disappear in FM1, while completely new dHvA
branches appear in the PM phase. Agreement between these experiments with the results of band calculations is not clear, because of the low symmetry
of the crystal structure and the large polarized moment. Nevertheless, a cylindrical shape of the Fermi surfaces is expected due to the flat
Brillouin zone along the $a$-axis. The cyclotron effective mass gradually increases with pressure. In the PM phase, a large mass ranging from $20$ to
$60\,m_0$ is detected, which is consistent with the pressure variation of the Sommerfeld coefficient $\gamma$ detected from specific heat
measurements.~\cite{Tateiwa2001}

\begin{figure}[tbv]
\begin{center}
\includegraphics[width=1 \hsize,clip]{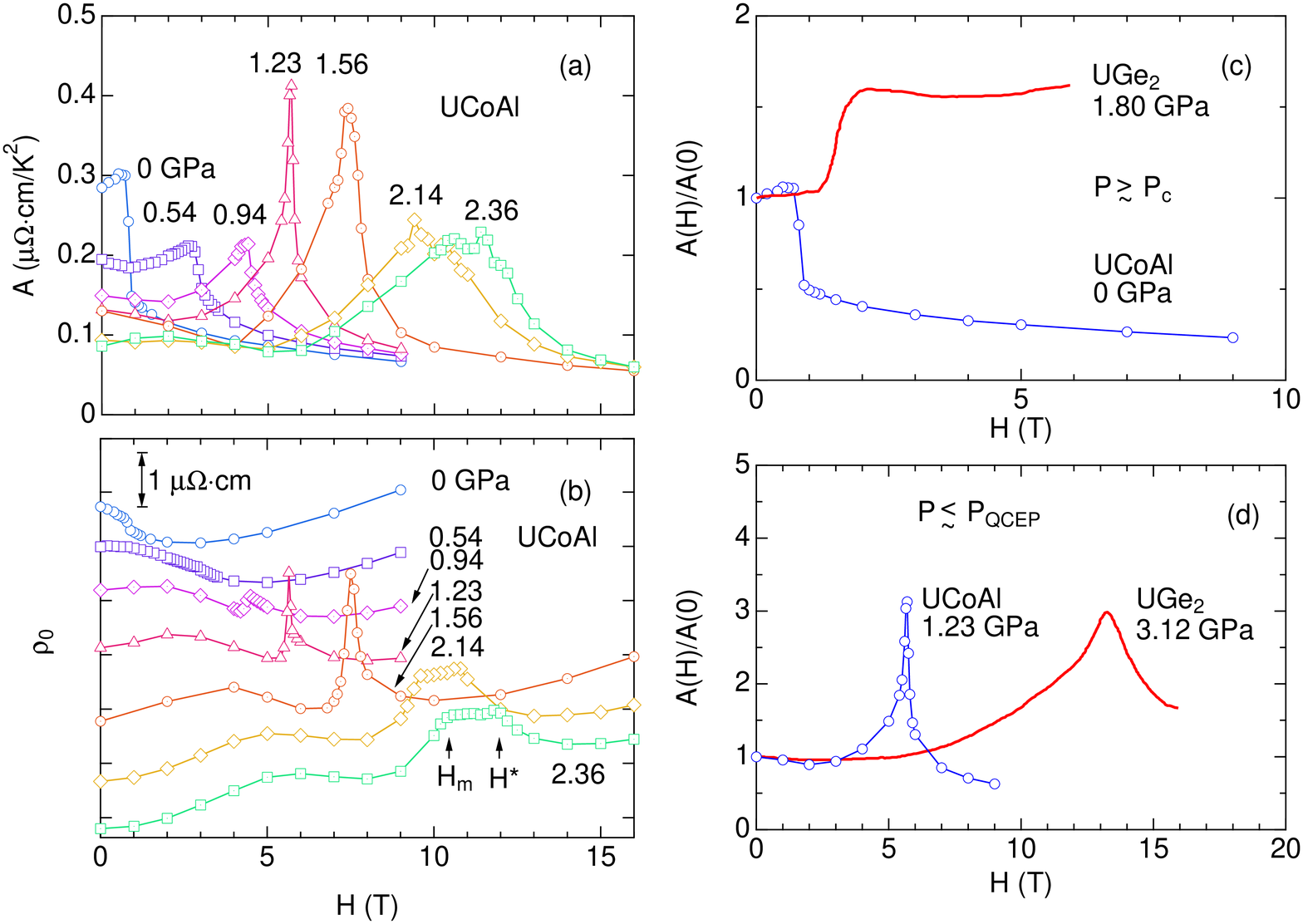}
\end{center}
\caption{Field dependence of the resistivity coefficient $A$ (a) and residual resistivity $\rho_0$ (b) at various pressures in UCoAl.
A comparison with the $A$ coefficient of UGe$_2$ is shown in panels (c) and (d)~\cite{Aoki2011a}.}
\label{fig:A_coef}
\end{figure}

In itinerant ferromagnets or nearly ferromagnets, a metamagnetic transition from PM to field-induced FM can occur. Well-known materials are the weak
ferromagnet ZrZn$_2$~\cite{Uhlarz2004} and the nearly ferromagnetic compound Sr$_3$Ru$_2$O$_7$~\cite{Grigera2001}. In particular, Sr$_3$Ru$_2$O$_7$
has attracted a strong interest following the finding of a new quantum phase called ``nematic'' phase~\cite{Borzi2007}. A difficulty to investigate
the QCEP in UGe$_2$ is to combine high magnetic field and high pressure, which prevents from a detailed study of ferromagnetic quantum criticality. A
new system which can be easily tuned to the ferromagnetic quantum criticality has been recently found: UCoAl~\cite{Aoki2011a}. As shown in
Fig.~\ref{fig:QCEP}(b), the ground state of UCoAl is PM, and a first-order metamagnetic transition from PM to FM is observed at low field for
$H\parallel c$. By applying pressure, the metamagnetic transition field $H_{\rm m}$ increases, and its first-order nature terminates at $P_{\rm
QCEP}\sim 1.5\,{\rm GPa}$ and $H_{\rm QCEP}\sim 7\,{\rm T}$, where a sharp enhancement of the effective mass is detected by resistivity measurements.
An interesting point is that the first-order metamagnetic transition changes into a second-order transition or crossover above the QCEP, and a new
phase appears between $H_{\rm m}$ and $H^\ast$, with a plateau in the resistivity coefficient $A$ and residual resistivity $\rho_0$ (see
Fig.~\ref{fig:A_coef}). More detailed studies focusing on the Fermi surface instabilities are now required.

\begin{figure}[tbv]
\begin{center}
\includegraphics[width=1 \hsize,clip]{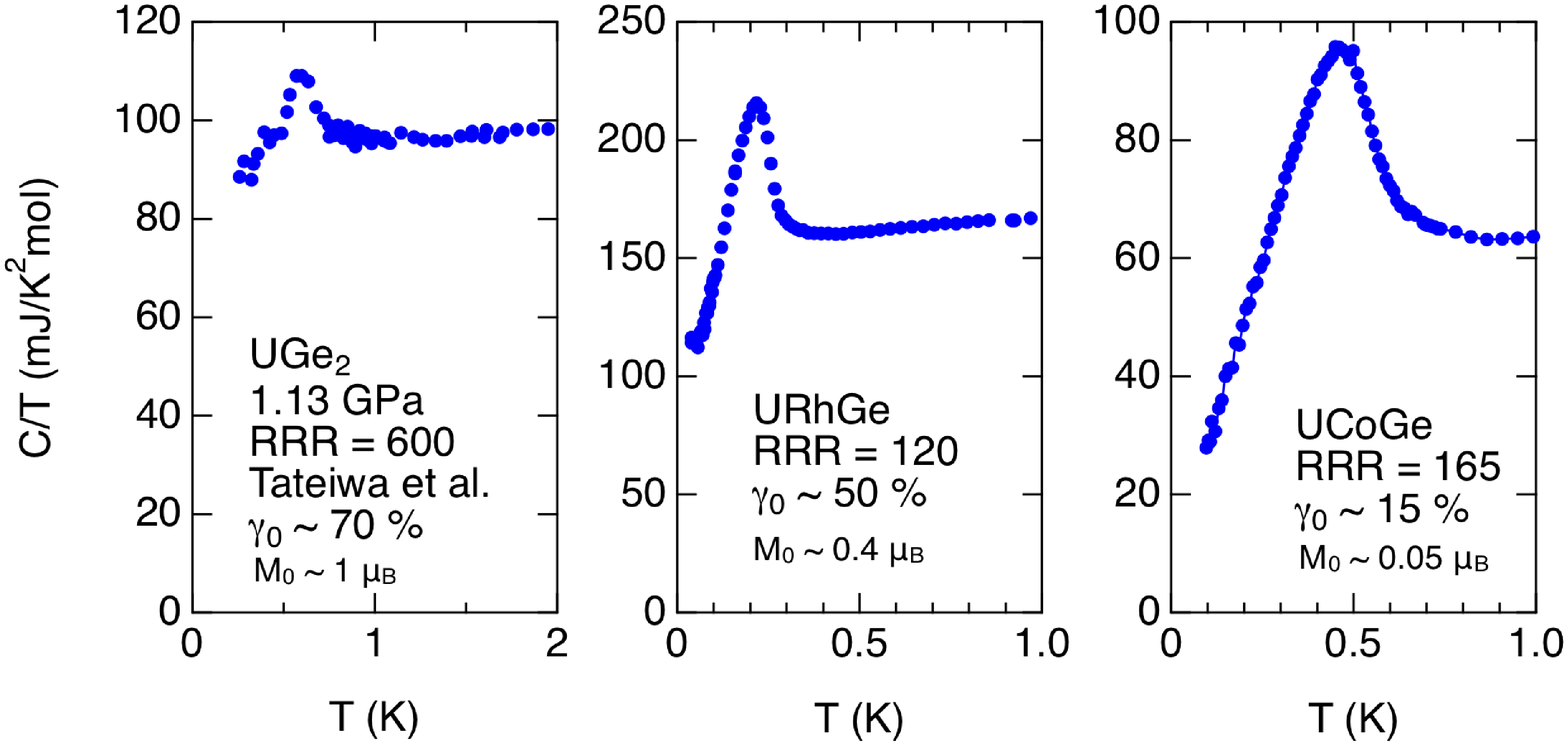}
\end{center}
\caption{Specific heat at low temperatures in UGe$_2$, URhGe and UCoGe~\cite{Tateiwa2001,Aoki2001,Aoki2012a}}
\label{fig:Cp}
\end{figure}

\begin{figure}[tbv]
\begin{center}
\includegraphics[width=.5 \hsize,clip]{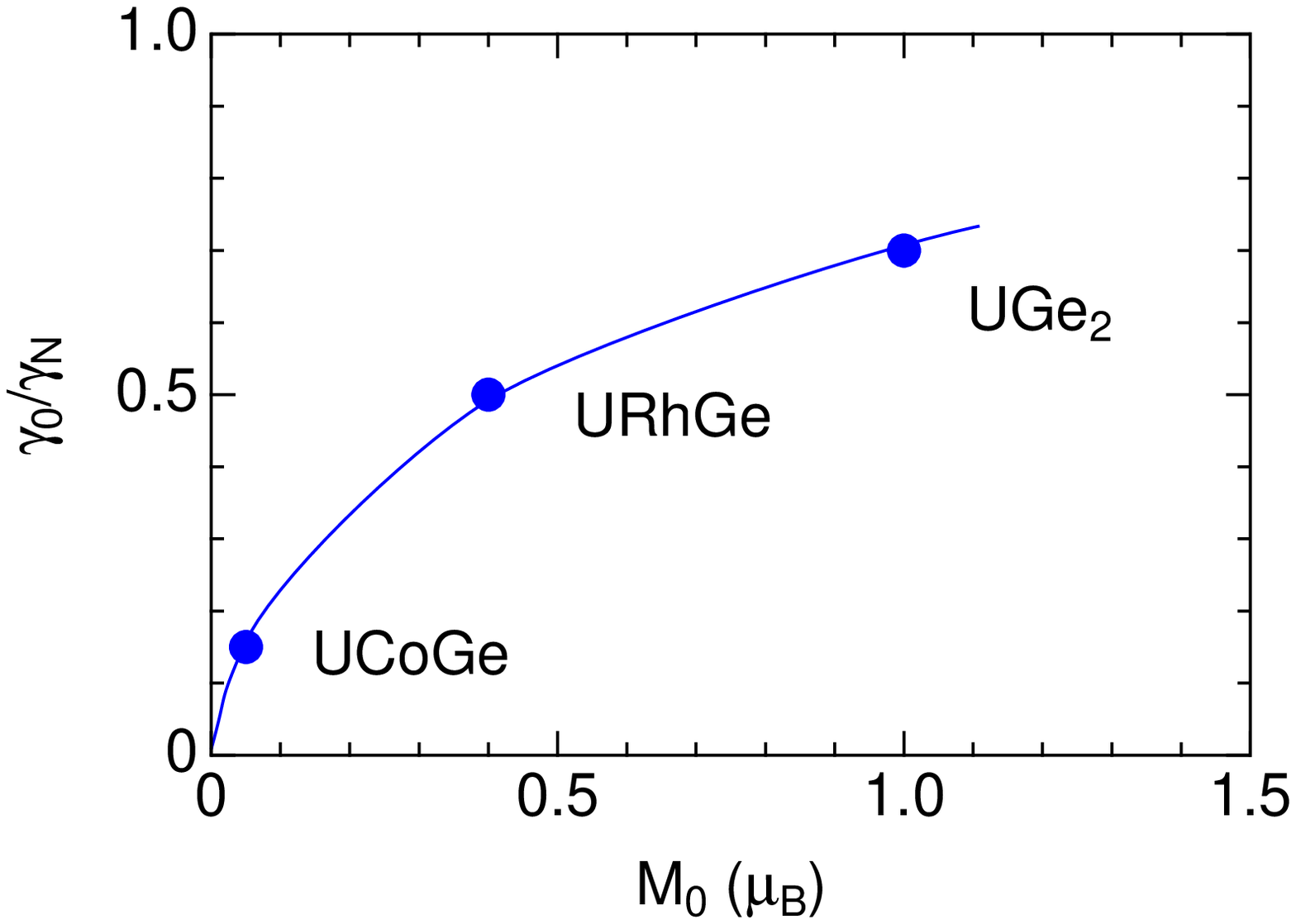}
\end{center}
\caption{Residual $\gamma$-value as a function of the ordered moment.}
\label{fig:gamma0}
\end{figure}

Evidences for the coexistence of FM and SC in UGe$_2$ were obtained by neutron diffraction~\cite{Huxley2001} and NQR experiments~\cite{Kotegawa2005}.
Bulk SC was confirmed by specific heat measurements, as shown in Fig.~\ref{fig:Cp}(a)~\cite{Tateiwa2001}. An interesting point is that a large
residual $\gamma$-term remains at $0\,{\rm K}$. For comparison, the specific heat results in URhGe and UCoGe are also shown in
Fig.~\ref{fig:Cp}(b-c)~\cite{Aoki2001,Aoki2012a}. Even for high-quality samples, as confirmed by their high residual resistivity ratio (RRR), the
residual $\gamma$-values ($\gamma_0$) are large, implying that a larger ordered moment yields a larger residual $\gamma$-value (see
Fig.~\ref{fig:gamma0}). A possible reason might be that one of the spin-components for the equal spin pairing is not gapped
($\Delta_{\downarrow\downarrow}\approx 0$), and the other spin-components are only gapped ($\Delta_{\uparrow\uparrow} \neq 0$) and are responsible
for the development of SC. Another possible reason might be that the materials are always in a SC mixed state (with no lower critical field $H_{\rm
c1}$), because of a large internal field proportional to the ordered moment.

\begin{figure}[bbv]
\begin{center}
\includegraphics[width=1 \hsize,clip]{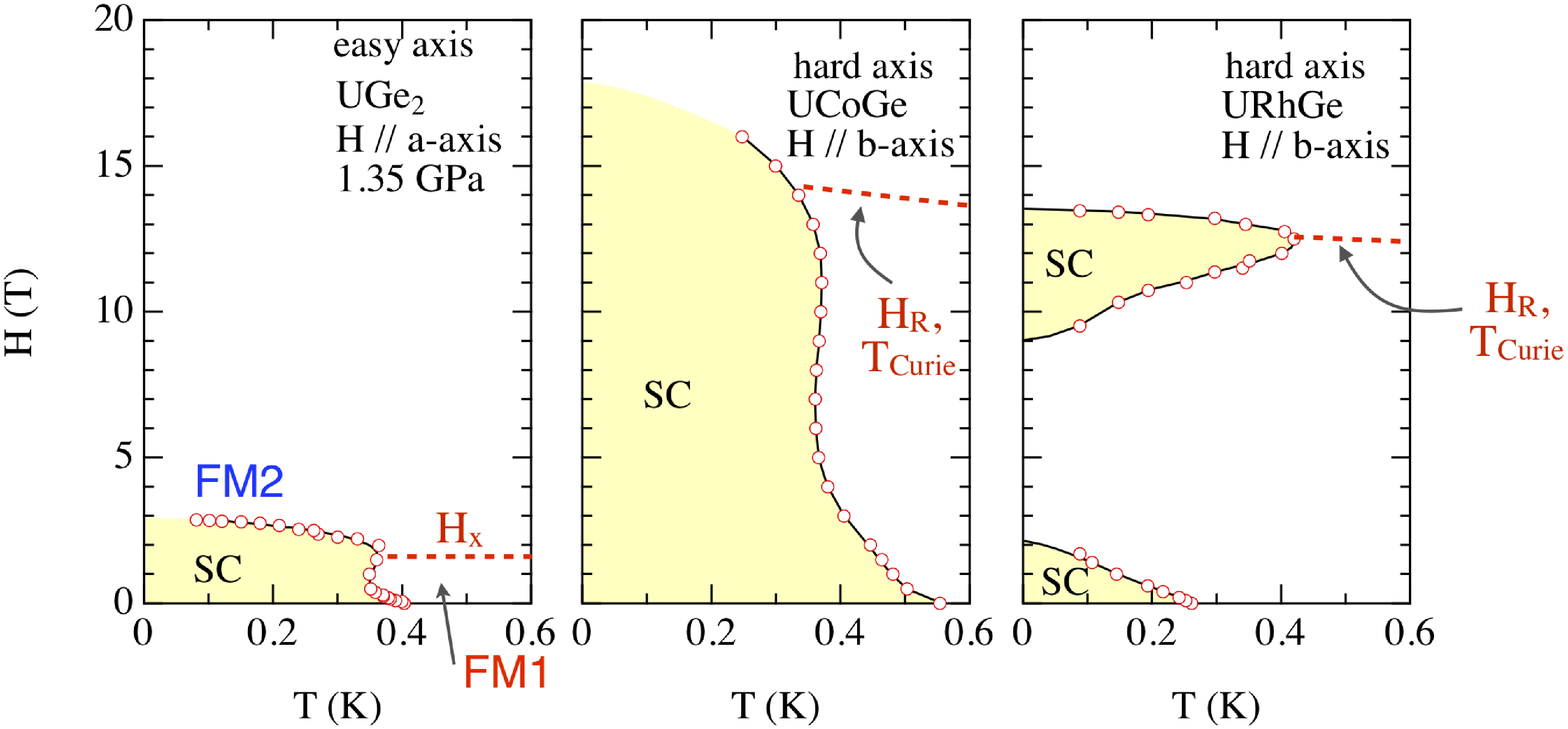}
\end{center}
\caption{$H_{\rm c2}$ curves of (a) UGe$_2$, (b) UCoGe and (c) URhGe.}
\label{fig:Hc2}
\end{figure}

When the pressure is tuned between $P_{\rm x}$ and $P_{\rm c}$, the upper critical field $H_{\rm c2}$ for $H\parallel a$ shows a peculiar temperature
dependence, as shown in Fig.~\ref{fig:Hc2}(a)~\cite{Sheikin2001a}. The S-shaped curve of $H_{\rm c2}$ is connected to the boundary between the two
different ferromagnetic states FM1 and FM2. The Fermi surface instabilities associated with the enhancement of the effective mass reinforce the SC
phase. The value of $H_{\rm c2}(0)$ is much higher than the Pauli limit expected for the weak coupling BCS scheme. Thus, it is natural that the
spin-triplet state with equal-spin paring is realized, where SC persists even in a large ferromagnetic internal field.

\subsection{Huge and anisotropic $H_{\rm c2}$ in URhGe and UCoGe}

URhGe shows SC at $T_{\rm sc}=0.25\,{\rm K}$ already at ambient pressure~\cite{Aoki2001}. $T_{\rm Curie}$ ($=9.5\,{\rm K}$) is much higher than
$T_{\rm sc}$, and the moment is directed along $c$-axis in the TiNiSi-type orthorhombic structure. The ordered moment is $0.4\,\mu_{\rm B}$ which is
considerably reduced from the free ion value, $3.6\,\mu_{\rm B}$, for both $5f^2$ and $5f^3$ configurations. The large specific heat $\gamma$-value
$160\,{\rm mJ/K^2 mol}$ indicates that URhGe is a heavy fermion compound. By applying pressure, $T_{\rm Curie}$ increases while $T_{\rm sc}$
decreases, indicating that pressure drives URhGe away from its critical region [see Fig.~\ref{fig:TP_phase}(b)], which can be also inferred from
thermal expansion measurements via applying the Ehrenfest relation~\cite{Hardy2005,Miyake2009,Aoki2011b}.

\begin{figure}[tbh]
\begin{center}
\includegraphics[width=0.5 \hsize,clip]{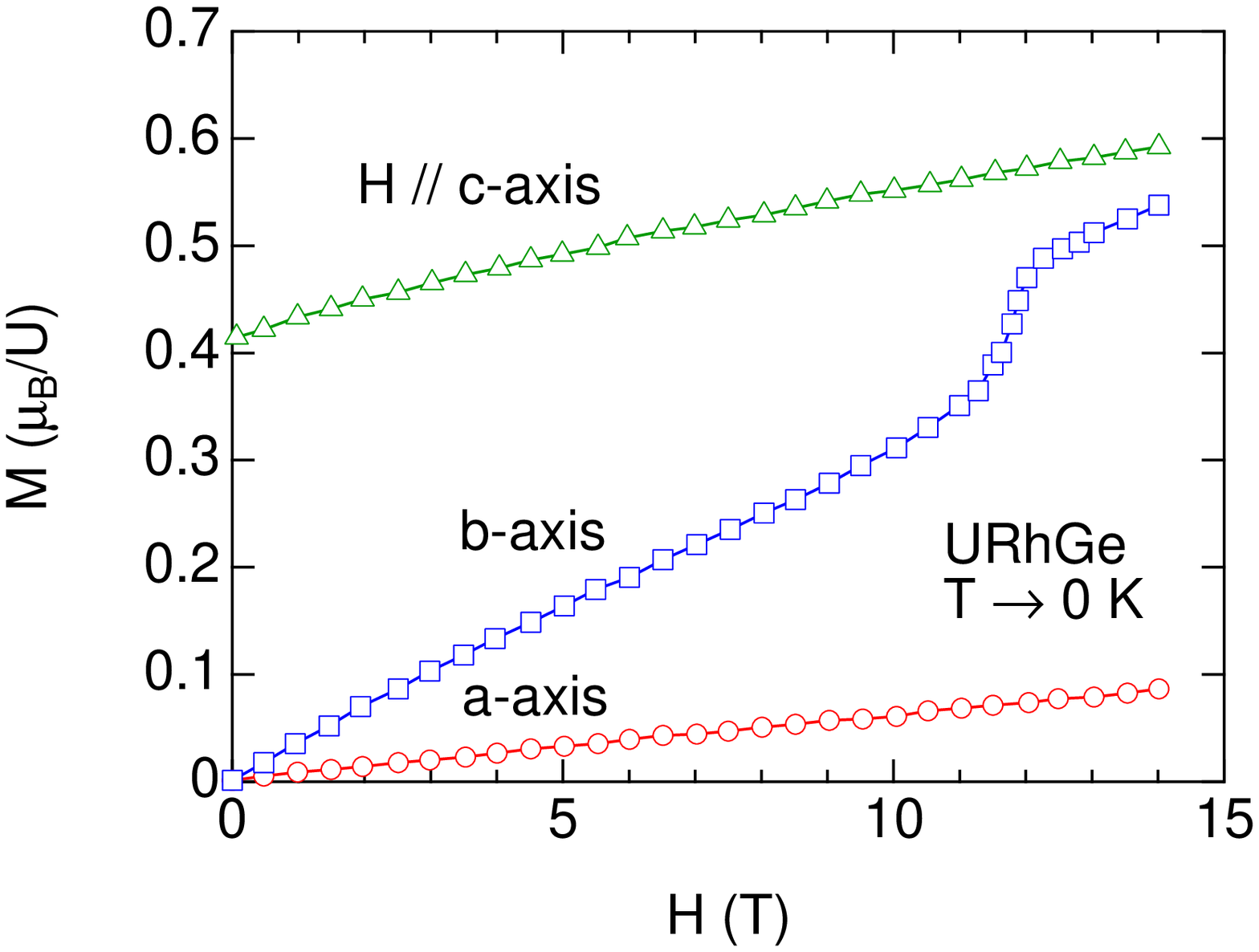}
\end{center}
\caption{Magnetization curves of URhGe.~\cite{Hardy2011}}
\label{fig:mag}
\end{figure}

Figure~\ref{fig:mag} shows the magnetization $M$ versus field $H$ curves extrapolated to $0\,{\rm K}$ for a magnetic field applied along the $a$, $b$
and $c$-axes~\cite{Hardy2011}. The initial slope of $M(H)$ for $H\parallel b$ is large compared to that of for $H$ along the easy-magnetization
$c$-axis, indicating that the moment starts to tilt from $c$ to $b$-axis, which finally ends by a spin reorientation occuring at $H_{\rm R}\approx
12\,{\rm T}$, where the moment is completely directed along the hard-magnetization $b$-axis.

\begin{figure}[tbh]
\begin{center}
\includegraphics[width=1 \hsize,clip]{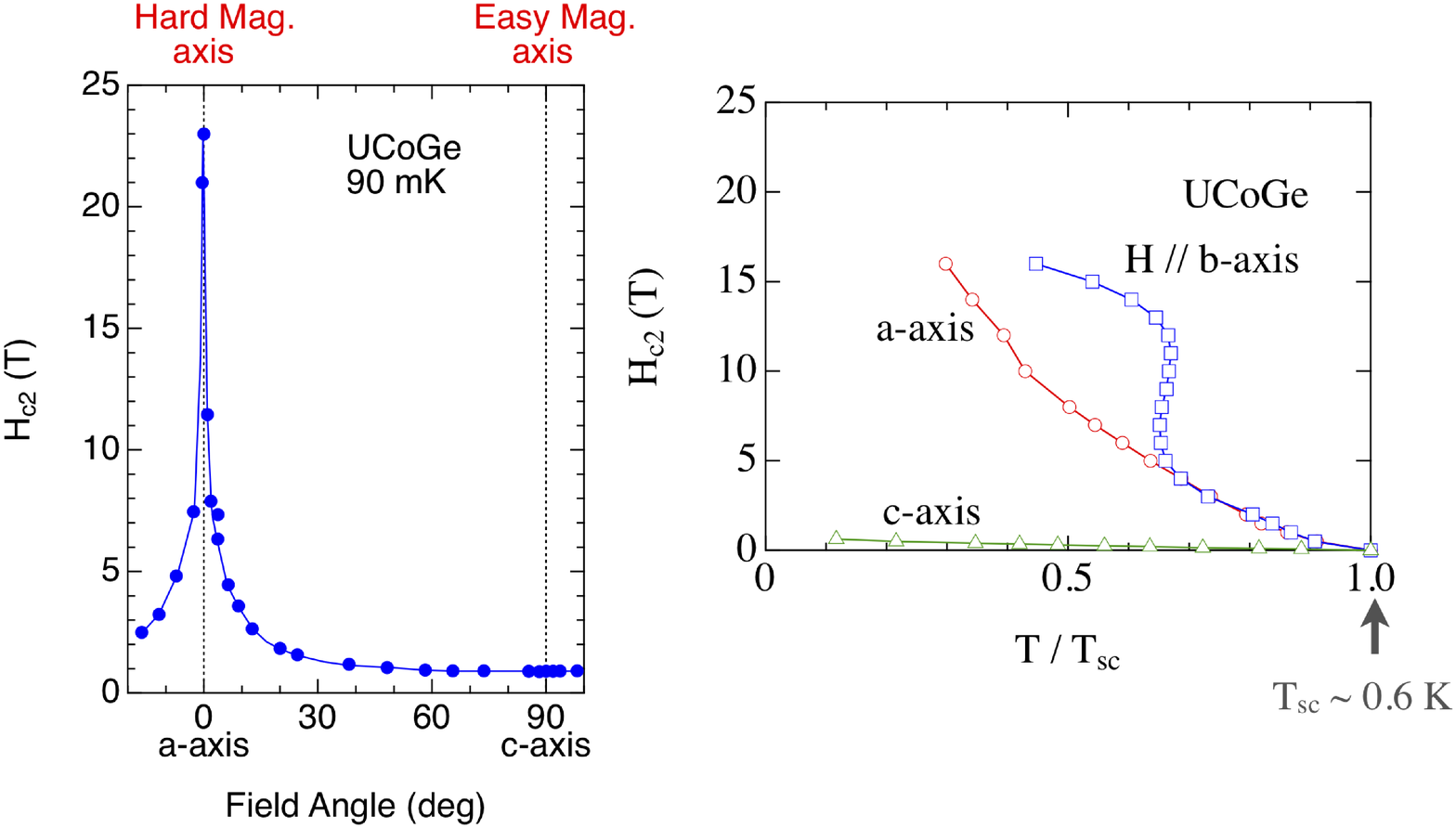}
\end{center}
\caption{(a) Angular dependence of $H_{\rm c2}$ at $90\,{\rm mK}$ for $H\parallel a$ $\to$ $c$, and (b) Temperature dependence of $H_{\rm c2}$ for $H
\parallel a$, $b$ and $c$.~\cite{Aoki2009}} \label{fig:UCoGe_Hc2}
\end{figure}

Surprisingly, field-reentrant superconductivity is found just below $H_{\rm R}$ for $H\parallel b$-axis~\cite{Levy2005}. As shown in
Fig.~\ref{fig:Hc2}(c), SC is first destroyed at $2\,{\rm T}$ but, in further increasing field, SC reappears in the field range from $8$ to $13\,{\rm
T}$. It should be noted that the critical temperature of SC at $12\,{\rm T}$ is higher than that at zero field, meaning that SC is indeed reinforced
under magnetic field. Since the reentrant SC dome at high fields is connected to the spin-reorientation, it is natural to consider that the
ferromagnetic fluctuations associated with the spin reorientation induce SC. If the field direction is slightly tilted from $b$ to $c$-axis, the
reentrant SC phase is immediately suppressed~\cite{Aoki2011c}. On the other hand, when the field direction is tilted from $b$ to $a$-axis, the
reentrant SC phase shifts to the higher field-range, following the $1/\cos\theta$ dependence of $H_{\rm R}$ ($\theta$: field angle from $b$ to
$a$-axis)~\cite{Levy2007}.

Similar field-reinforced SC is also found in UCoGe~\cite{Aoki2009}. The crystal structure of UCoGe is identical to that of URhGe. In UCoGe, $T_{\rm
Curie}$ is equal to $2.5\,{\rm K}$ and SC is observed at $0.6\,{\rm K}$ at ambient pressure~\cite{Huy2007}. As shown in Fig.~\ref{fig:TP_phase}(c),
$T_{\rm Curie}$ decreases with increasing pressure and reaches zero at $P_{\rm c}\sim 1\,{\rm GPa}$, while $T_{\rm sc}$ increases, shows a maximum at
$P_{\rm c}$, and then decreases, indicating that SC survives even in the PM phase~\cite{Hassinger2008,Slooten2009}. This phase diagram is quite
different from that obtained in UGe$_2$ and also from theoretical predictions that $T_{\rm sc}$ should drop at $P_{\rm c}$.

$H_{\rm c2}$ in UCoGe is quite anisotropic. As shown in Fig~\ref{fig:UCoGe_Hc2}, $H_{\rm c2}$ for $H \parallel c$ is rather small, since it equals
$\sim 1\,{\rm T}$, which is close to the Pauli limit. However, when the field is applied along the hard-magnetization $a$ and $b$-axes, a huge
$H_{\rm c2}$ is observed~\cite{Aoki2009}. For $H\parallel a$, the $H_{\rm c2}$ curve shows an unusual upward curvature, $H_{\rm c2}$ probably
reaching more than $25\,{\rm T}$ at $0\,{\rm K}$. For $H\parallel b$, an unusual $H_{\rm c2}$ curve with a ``S''-shape is further observed, $H_{\rm
c2}$ at $0\,{\rm K}$ being also huge ($\sim 20\,{\rm T}$). When the field is slightly tilted to the easy-magnetization $c$-axis, $H_{\rm c2}$ at low
temperature is strongly suppressed as shown in Fig.~\ref{fig:UCoGe_Hc2}. The values of $H_{\rm c2}(0)$ for $H\parallel a$ and $b$ greatly exceed the
Pauli limit, indicating that a spin-triplet state with equal spin pairing is responsible for SC.

\begin{figure}[tbv]
\begin{center}
\includegraphics[width=1 \hsize,clip]{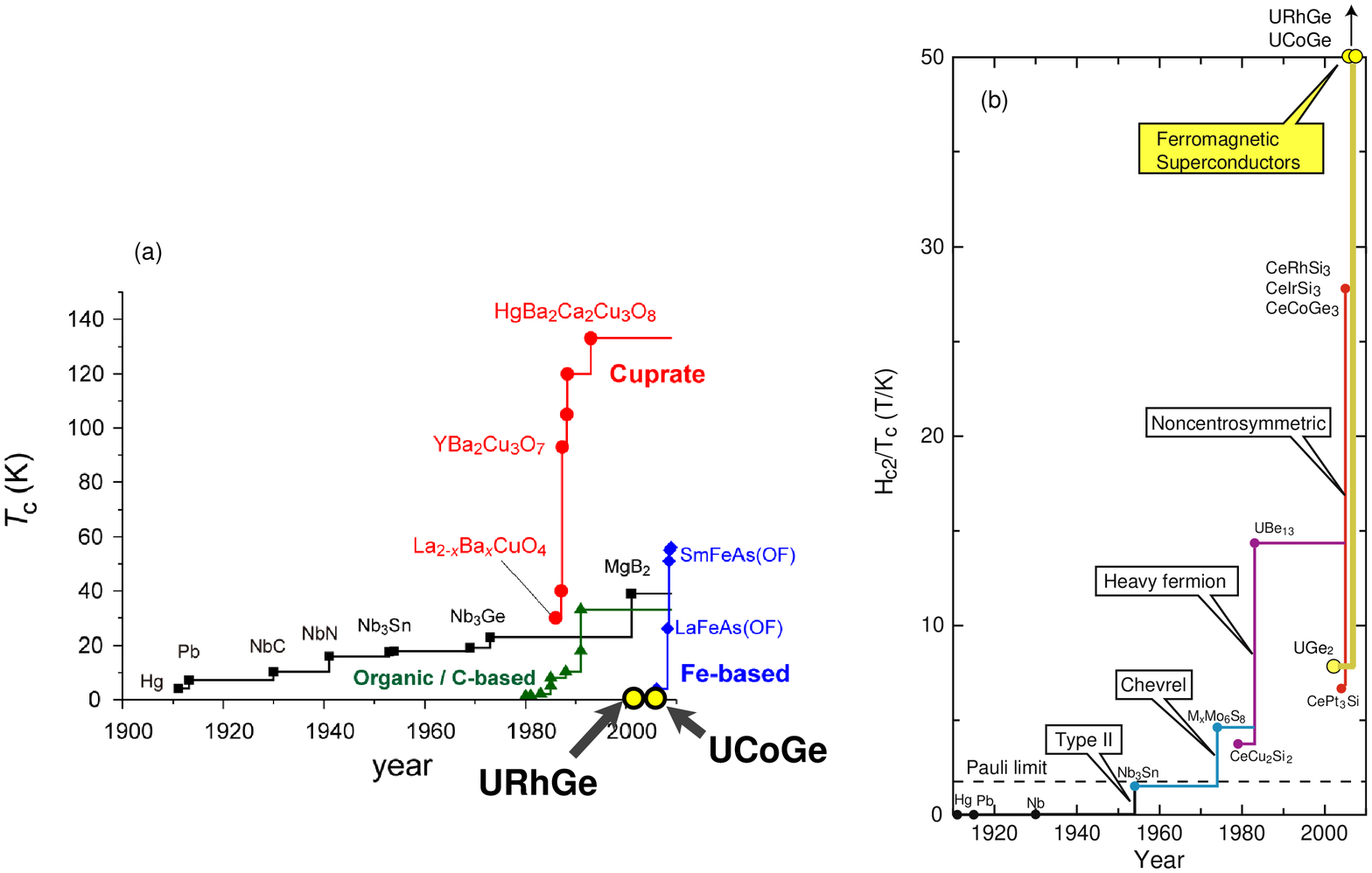}
\end{center}
\caption{History of $T_{\rm sc}$ and $H_{\rm c2}/T_{\rm sc}$~\cite{Kittaka,Kimura2010}.}
\label{fig:Hc2_year}
\end{figure}

Fig.~\ref{fig:Hc2_year} (a) shows that the highest superconducting temperature has been strongly enhanced since 100 years~\cite{Kittaka}, from
conventional to unconventional superconductors, including the recently discovered new class of U-based superconductors. Fig.~\ref{fig:Hc2_year} (b)
shows the evolution with time of $H_{\rm c2}$ divided by $T_{\rm sc}$ for various superconductors ~\cite{Kimura2010}. In general $H_{\rm c2}$ is
governed either by the Pauli limit or by the orbital limit. In conventional BCS superconductors, the initial slope of $H_{\rm c2}(T)$ at $H=0$ is
small and $H_{\rm c2}$ at $0\,{\rm K}$ is also small. In heavy-fermion superconductors, the initial slope is potentially large because of the large
effective mass. $H_{\rm c2}$ is, however, limited by the Pauli limit due to the Zeeman splitting based on the spin-singlet state. In the case of no
Pauli paramagnetic effect due to a spin-triplet state, $H_{\rm c2}$ would be only governed by the orbital limit. Thus, a huge $H_{\rm c2}$ is
expected in heavy fermion compounds, where the orbital limit can be described as $\Psi/2\pi\xi^2$ or $0.73\, dH_{\rm c2}/dT\, T_{\rm sc}$ within the
clean limit. This is indeed realized in non-centrosymmetric compounds and ferromagnetic superconductors.

\begin{figure}[tbv]
\begin{center}
\includegraphics[width=1 \hsize,clip]{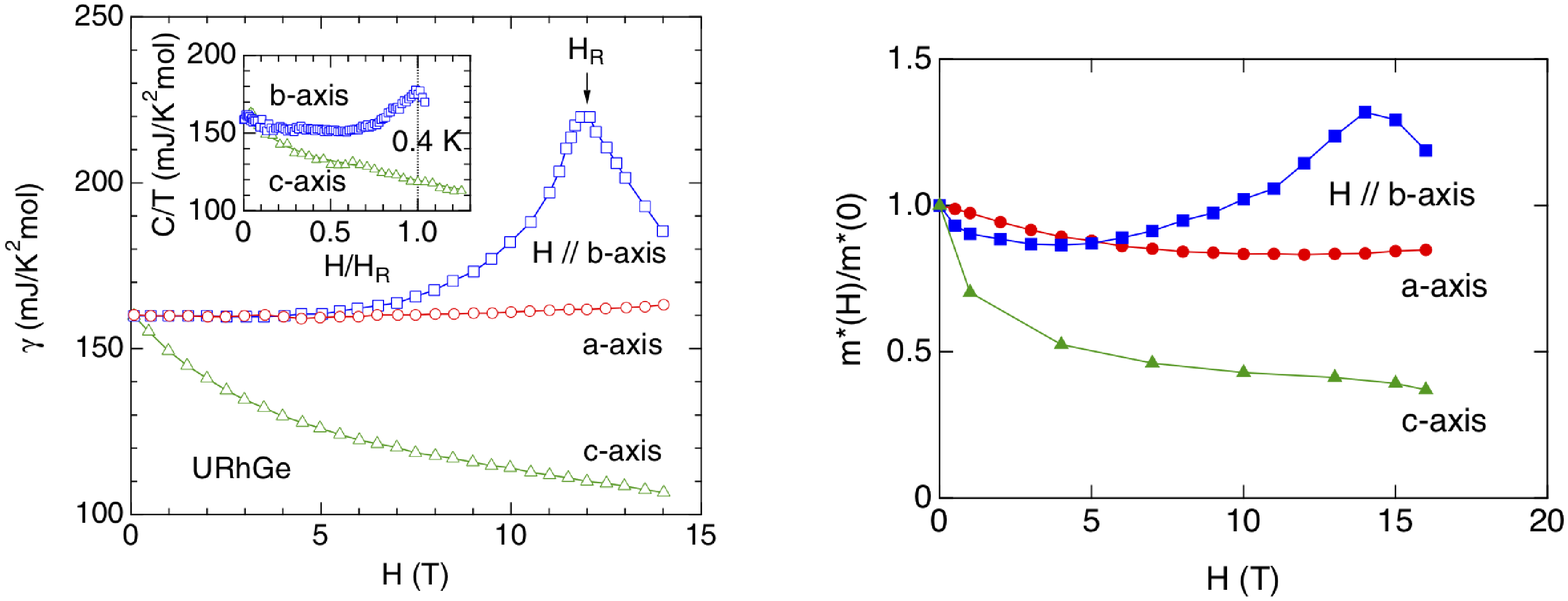}
\end{center}
\caption{(a) Field dependence of the specific heat $\gamma$-value for $H\parallel a$, $b$ and $c$ in URhGe~\cite{Hardy2011} and (b) field dependence
of the resistivity $A$ coefficient in UCoGe~\cite{Aoki2009}. The data in the main panel of (a) were obtained from the temperature dependence of the
magnetization using the Maxwell relation. The inset in (a) shows direct specific heat measurements.} \label{fig:gamma}
\end{figure}

As shown in Fig.~\ref{fig:gamma}, the field response of the specific heat $\gamma$-value is anisotropic both in URhGe and in
UCoGe~\cite{Miyake2008,Hardy2011,Aoki2009,Aoki2011c}. The $\gamma$-value for $H\parallel c$ decreases with increasing field, as expected for usual
weak ferromagnets where the spin fluctuations are suppressed under magnetic field. Oppositely, for $H\parallel b$ $\gamma$ shows a maximum at the
field where the field-reinforced SC (or RSC) is observed, indicating that ferromagnetic fluctuations are enhanced. For $H \parallel a$, the
$\gamma$-value retains a large value, but a maximum is not observed.

In a simple model based on a McMillan-like formula, $T_{\rm sc}$ can be written as $T_{\rm sc}\sim \exp(-m^\ast/m^{\ast\ast})$, where the effective
mass $m^\ast$ can be described by the band mass $m_{\rm B}$ and the extra mass $m^{\ast\ast}$ is directly related to the SC pairing, with namely
$m^\ast = m_{\rm B}+m^{\ast\ast}$. If the extra mass (correlation mass) $m^{\ast\ast}$ is enhanced, $T_{\rm sc}$ is also enhanced. Furthermore, as
$H_{\rm c2}$ is governed by the orbital limit ($\sim \psi_0/2\pi \xi^2$), $H_{\rm c2}$ can be simply described as $H_{\rm c2}\propto (m^\ast T_{\rm
sc})^2$, using the relations $\xi \sim \hbar v_{\rm F}/k_{\rm B}T_{\rm sc}$ and $m^\ast v_{\rm F} = \hbar k_{\rm F}$. Therefore, if the effective
mass is enhanced, a large $H_{\rm c2}$ can be expected. It is worth noting that here we simply assume invariant Fermi surfaces under magnetic field.

\begin{figure}[tbv]
\begin{center}
\includegraphics[width=1 \hsize,clip]{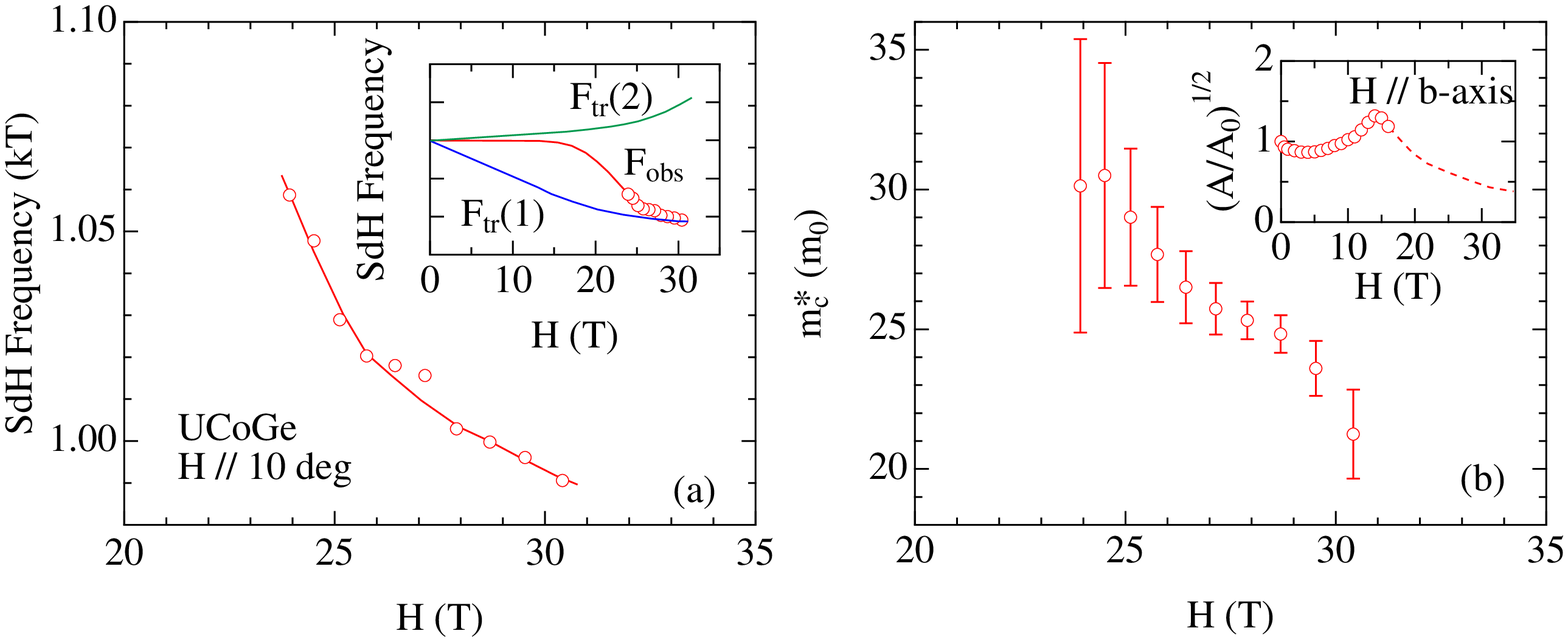}
\end{center}
\caption{(a) Field dependence of the SdH frequency at high field in UCoGe and (b) the field dependence of the cyclotron effective mass.}
\label{fig:UCoGe_SdH}
\end{figure}

An interesting question is whether the Fermi surface is really unchanged near the field-reinforced SC region. In UCoGe, Shubnikov de Haas (SdH)
experiments~\cite{Aoki2011} show that the observed SdH frequency ($\sim 1\,{\rm kT}$) decreases with increasing field just above $H_{\rm c2}$ (see
Fig.~\ref{fig:UCoGe_SdH}), indicating a non-linear field response of the Zeeman-split Fermi surface. Correspondingly, the cyclotron mass also
decreases with increasing field. A shrinkage of the Fermi surface near RSC was reported from SdH experiments in URhGe.~\cite{Yelland2011} Since the
Fermi surfaces observed in these experiments carry only $12\,\%$ and $1.5\,\%$ of the total specific heat $\gamma$-value in UCoGe and URhGe,
respectively, a definite conclusion is still under debate. It should be noted that recent high-field thermopower measurements on UCoGe show a sharp
peak around $11\,{\rm T}$ for $H\parallel b$, suggesting the modification of the Fermi surfaces~\cite{Malone2012}.

In summary of this section, high-field experiments have revealed a new aspect of ferromagnetic quantum critical phenomena and field-reinforced SC in
ferromagnetic superconductors. A key point is the Ising-type of the longitudinal ferromagnetic fluctuations, which induces an anisotropic and huge
$H_{\rm c2}$. On the other hand, it has been recently proposed that XY-type transverse magnetic fluctuations might play an important role to raise
$T_{\rm sc}$ in AF or nearly AF compounds. Precise experiments with microscopic experimental probes, such as NMR, neutron scattering and quantum
oscillations, are required for the future studies.


\section*{Acknowledgments}

We acknowledge G. Ballon, F. Bourdarot, D. Braithwaite, J. P Brison, A. Buzdin, T. Combier, A. Demuer, J. Flouquet, S. Fujimoto, K. Goetze, P. Haen,
F. Hardy, H. Harima, K. Hasselbach, E. Hassinger, T. Hattori, L. Howald, A. Huxley, D. Hykel, K. Ishida, N. Kimura, G. Knebel, H. Kotegawa, G.
Lapertot, P. Lejay, F. L\'{e}vy, L. Malone, T.D. Matsuda, C. Meingast, V. Michal, V. Mineev, A. Miyake, K. Miyake, Y. \={O}nuki, C. Paulsen, C.
Proust, S. Raymond, G. Scheerer, R. Settai, J. Spalek, Y. Tada, V. Taufour, D. Vignolle, A. de Visser, S. Watanabe, J. Wosnitza, H. Yamagami, and E.
Yelland. This work was supported by the French ANR DELICE, by Euromagnet II via the EU under Contract No. RII3-CT-2004-506239, and by the ERC
Starting Grant NewHeavyFermion.

\bibliographystyle{ieeetr}
\bibliography{cras}

\end{document}